\documentclass[12pt,preprint]{aastex}

\usepackage{graphicx}

\slugcomment{Not to appear in Nonlearned J., 45.}

\shorttitle{Quasi-periodicities in Blazars}
\shortauthors{A. Sandrinelli et al.}

\begin{document}


\title{Quasi-periodicities at year-like timescales in Blazars}


\author{A. Sandrinelli} 
\affil{
\textit{Universit\`a degli Studi dell'Insubria, Dipartimento di Scienza ed Alta Tecnologia, \\ Via Valleggio 11, I-22100 Como, Italy.\\
INAF - Istituto Nazionale di Astrofisica, Osservatorio Astronomico di Brera, \\Via Emilio Bianchi 46, I-23807 Merate, Italy.}
    }

\author{S. Covino}
\affil{\textit{
INAF - Istituto Nazionale di Astrofisica, Osservatorio Astronomico di Brera, \\Via Emilio Bianchi 46, I-23807 Merate, Italy
}}

\author{M. Dotti}
\affil{\textit{
Universit\`a degli Studi di  Milano Bicocca, 
Dipartimento di Fisica G. Occhialini,\\ Piazza della Scienza 3, I-20126 Milano, Italy\\
INFN- Istituto Nazionale di Fisica Nucleare, \\Universit\`a degli Studi di  Milano Bicocca, 
Dipartimento di Fisica G. Occhialini,\\ Piazza della Scienza 3, I-20126 Milano, Italy
}}

\and

\author{A. Treves\altaffilmark{1}}
\affil{\textit{Universit\`a degli Studi dell'Insubria, Dipartimento di Scienza ed Alta Tecnologia,\\ Via Valleggio 11, I-22100 Como, Italy\\
INAF - Istituto Nazionale di Astrofisica, Osservatorio Astronomico di Brera, \\Via Emilio Bianchi 46, I-23807 Merate, Italy\\
INFN- Istituto Nazionale di Fisica Nucleare,  Sezione Trieste - Udine, \\
 Via Valerio 2, I-34127 Trieste, Italy}
}

\altaffiltext{1}{ICRA}



\begin{abstract}
We searched for quasi-periodicities on year-like timescales in the light curves of 6
    blazars in the optical - near infrared bands  and 
    we made a comparison with the high energy emission.
We obtained optical/NIR light curves from REM photometry plus archival SMARTS 
 data and we accessed the \textit{Fermi} light curves for the $\gamma$-ray data. 
  The periodograms often show strong peaks
   in the optical and $\gamma$-ray bands,  which in some cases 
 may be inter-related. 
The significance of the  revealed peaks is then discussed,
  taking into account that the noise is frequency dependent.
  Quasi-periodicities on a year-like timescale appear to occur often in blazars.   
No straightforward model describing these possible periodicities
is yet available, but some plausible  
 interpretations for the physical mechanisms
 causing  periodic variabilities of these sources are examined.
\end{abstract}


\section{Introduction}

Blazars are active galactic nuclei exhibiting  large variability 
at all frequencies. 
 Their emission is dominated by a relativistic jet,  the amplification 
 factor of the intensity is characterized by a power of the
 Doppler factor \citep[e.g.][]{Ghisellini2013}.
 The jet produces a non thermal spectrum  where often one can distinguish
 a synchrotron and  a Compton component.  
The energy source is assumed to be a combination of accretion on a supermassive
black hole (SMBH) 
and extraction of its spin energy. In some cases the presence of an accretion
disk is suggested by the appearance in the spectral energy distribution 
 of a thermal component.
The observed variability is supposedly a consequence of the intrinsic change of the
accretion onto the SMBH, of the jet formation process, and specifically of its beaming.
 Various  processes  originating in different regions and with different time
 scales result in a complicated  variability 
  leading to rather chaotic light curves.
  
 The study of variability through auto- and cross-correlation procedures
has  proven to be effective in constraining complex models of these sources. 
The discovery of a periodicity in the variability could have
profound consequences in the global understanding of the sources,  constituting a
basic block for  models.  
The effort for finding periodicities has been
substantial at all frequencies \citep[see e.g.][]{Falomo2014}. A  rather robust claim
 is that  of a $\sim$12 year period in OJ 287 \citep[e.g.][]{Sillanpaa1988}, 
 a BL Lac object   possibly containing in its center a system of two SMBHs
 \citep{Lehto1996}. 
 Note however that the periodicity and the picture are disputed for instance by \cite{Hudec2013}.
  \cite{Graham2015}  proposed a 1980 day optical period for the quasar PG1302-102,
which also appears reliable.
Models based on the presence of a
 binary black hole \citep[e.g.][]{Sundelius1997,Kidger2000,Valtonen2012}, 
 describe outbursts and flares 
following  a not strictly periodic  cadence. 
In these cases, convolution of various processes can lead to an apparently
 unstable or variable period, i.e. to quasi-periodicities.
Search for quasi-periodicities in blazars is a valid and efficient diagnostic tool.
Some sources could exhibit long lasting quasi-periodic behaviors. The time scales 
of the events and their persistence in the light curves  may allows us to shed light on the physical
 processes underlying these variations.
Among previous  proposals of year-like quasi-periodicities
in blazars, we refer to \cite{Raiteri2001}, \cite{Gabanyi2007}, \cite{Rani2009}, 
\cite{Li2015}, and references therein.
  In addition, recently \cite{Zhang2014} collecting photometric data of PKS 2155-304
published in the last 35 years, discovered  a quasi-periodicity of $T_{0}= $ 317 d. 
This was confirmed by our independent photometry \citep{Sandrinelli2014a,
Sandrinelli2014b}.  
Moreover we showed that a periodicity appears also in the $\gamma$-rays 
observed by the \textit{Fermi} mission at $T=2\cdot T_{0}$.

Because \textit{Fermi} has monitored continuously the sky since $\sim$ 6 years, 
it is obvious that it is only now that one can combine optical and $\gamma$-ray 
searches for year-like  periodicities, a procedure which was 
 successful for PKS 2155-304. 
 Since the number of covered periods in the optical and in $\gamma$-rays 
is limited, as in the case of PKS 2155-304, 
it is difficult to assess the stability on a long-term basis of the inferred (quasi) periodicities.

The starting point of the present investigation are the VRIJHK photometric 
observations obtained with the robotic 
 Rapid Eye Mounting telescope  \citep[REM\footnote{REM data can be retrieved from
 \texttt{http://www.rem.inaf.it}},][]{Zerbi2004, Covino2004} 
at La Silla, which are described in detail in 
\cite{Sandrinelli2014a}.
 Among the blazar sources  monitored by REM and described in the above mentioned 
 paper  we consider here PKS 0537-441, 
 OJ 287,   PKS 1510-089, PKS 2005-489 and  
 PKS 2155-304, because of the  extensive coverage of the  observations. 
 These data are available on-line\footnote{Photometric nightly averaged data  
 is tabled in
\texttt{http://vizier.cfa.harvard.edu/viz-bin/VizieR?-source\\=J/A+A/562/A79}}.
 We add also the REM photometry of 3C 279, which is unpublished thus far.
We have then combined the REM data with those derived from the Small \& Moderate 
Aperture Research Telescope System 
archives \citep[SMARTS\footnote{\texttt{http://www.astro.yale.edu/smarts/glast/home.php}},][]{Bonning2012}. 
 The REM data on PKS 2155-304 were originally examined in  \cite{Sandrinelli2014b} 
and are those which led us to confirm the results of \cite{Zhang2014}.
In Table \ref{sample} we report a summary of the characteristics of the six sources, 
of which two are flat spectrum radio quasars (FSRQs)
and the others are BL Lac objects.

The unpublished photometry of  3C 279 is described in Section \ref{photometry}. 
The search for periodicities from both  optical data and 
\textit{Fermi} archives is presented  in Section \ref{search}.  
 Discussion of results  with a possible picture for the interpretation
of a $\sim$1 yr quasi-periodicity in blazars is reported in Section \ref{disc}
and conclusions in Section \ref{concl}.


\section{\label{photometry}REM photometry of 3C 279}

 
The analysis of the data follows closely the scheme described by
\cite{Sandrinelli2014a}.  For the optical bands we  used reference 
stars   from \cite{Gonzalez2001},  while for the NIR frames we referred to
the \textit{Two Micron All Sky Survey 
Catalog }
\citep[2MASS\footnote{\texttt{http://www.ipac.caltech.edu/2mass/}},][]{Skrutskie2006}.

\begin{figure*}
\centering
\includegraphics[trim=0cm 2.5cm 0cm 0cm,clip,width=0.7\textwidth]{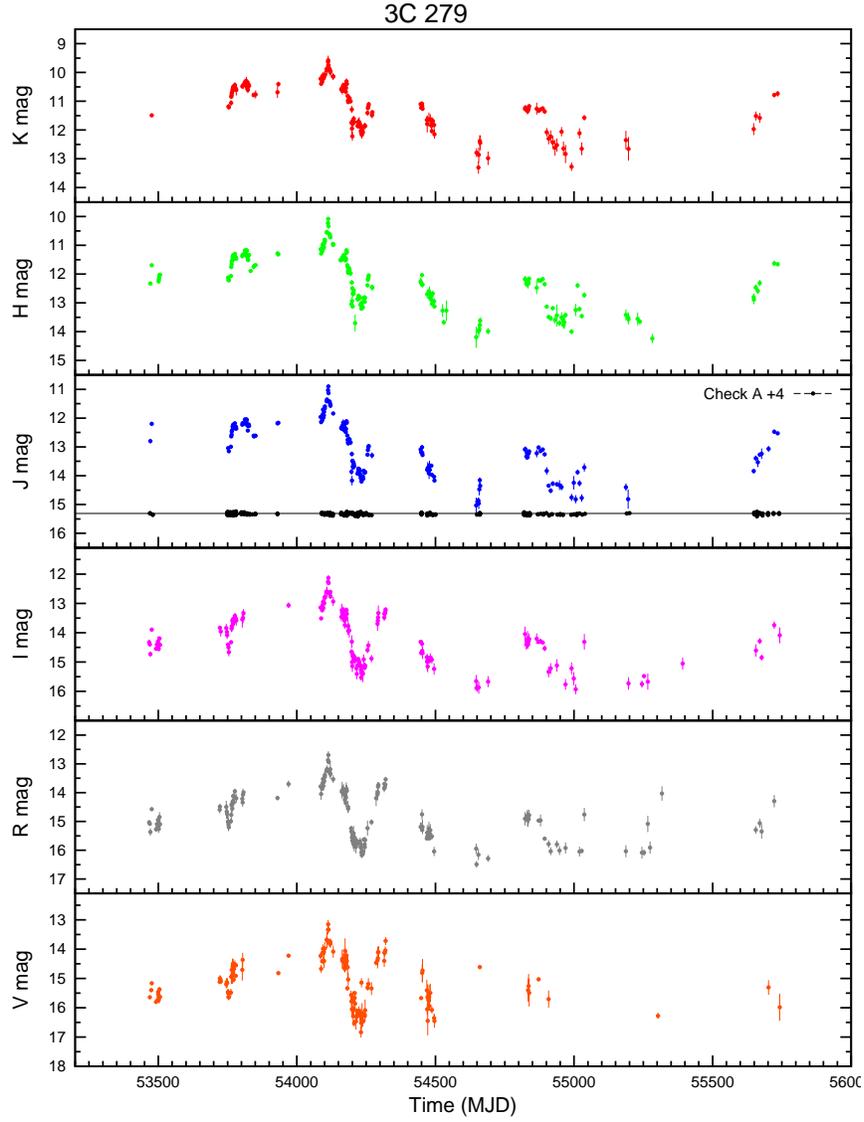}
\caption
{\label{lc} REM near-infrared and optical nightly averaged light curves of 3C 279.
The light curve of the check star is reported in J band (black points) with the indicated 
displacements $\Delta$m.}
\end{figure*}

 All images have been visually  checked
eliminating those where the targets or the reference stars are  
close to the borders of the frame, and where obvious  biases 
 were present.  In Table \ref{3c279lc_tab}  we report our photometry of the source. 
Some overall properties are given in Table \ref{3c279prop_tab}. 
Comparing the  fractional variability amplitude  $\sigma_{rms}$ 
 \citep[e.g.][]{Nandra1997,Edelson2002} 
   with those of the other 5 objects, 3C 279 appears highly variable. 
The nightly averaged light curves in the 6 filters  
are reported in Figure \ref{lc}.


  \section{\label{search}Search for periodicities in the target sources}
  
\begin{figure*}
\centering 
\includegraphics[trim=0cm 1cm 0cm 0cm,clip,width=.8\textwidth]{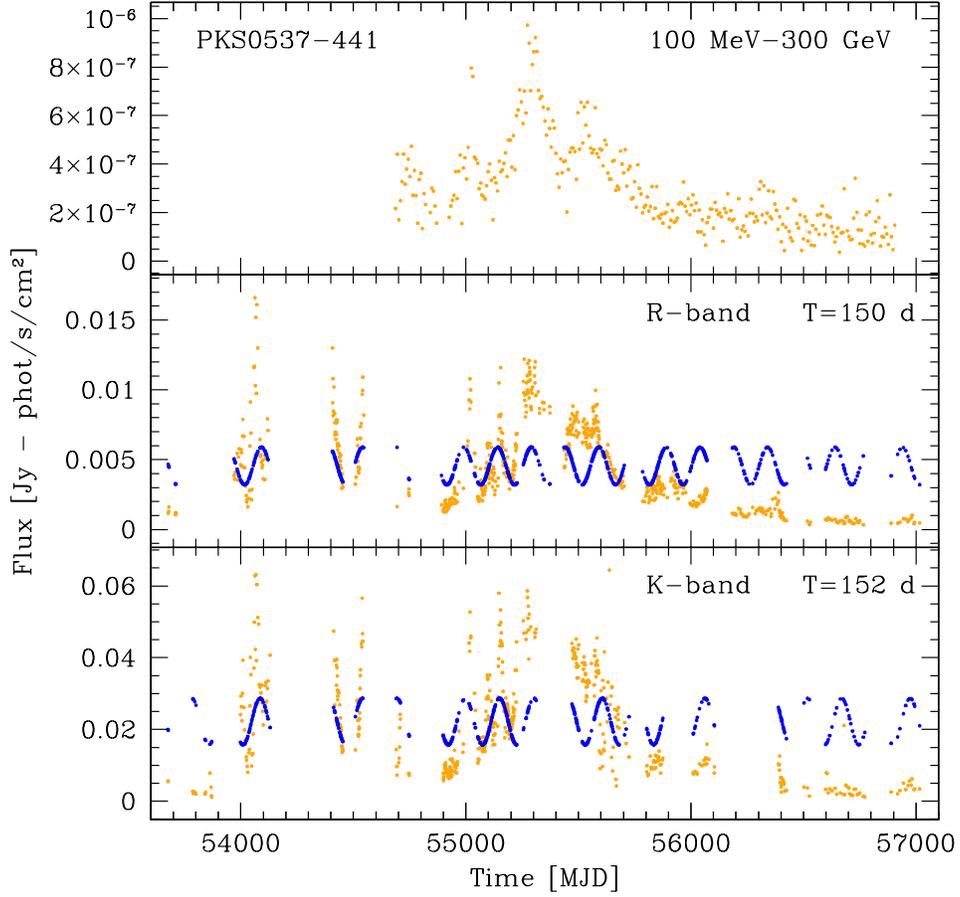}
	\caption{ 
	\label{3lc_pks0537}  
        Weekly averaged {Fermi} $\gamma$-ray  light curve in the 100 
	MeV - 300 GeV  energy range (\textit{top panel}, yellow points).
	Nightly averaged REM and SMARTS  light curves in R and  K bands
	are also reported 
	(\textit{central and bottom panels}, yellow points).
	Flux is 
	in photons $ \cdot $ s$^{-1}\cdot$ cm$^{2}$
	for the $\gamma$-ray light  curve and in Jy for the NIR-optical light curves.
	Errors are omitted for readability.
	The blue lines are the sinusoidal artificial models (see text).
	The amplitudes  are 
	A=0.001346 Jy  in R  (T=150 d), \ 
	and A=0.00647 Jy in K band  (T=152  d).			
	}
\end{figure*}
\begin{figure*}
\centering 
\includegraphics[trim=0cm 1cm 0cm 0cm,clip,width=.8\textwidth]{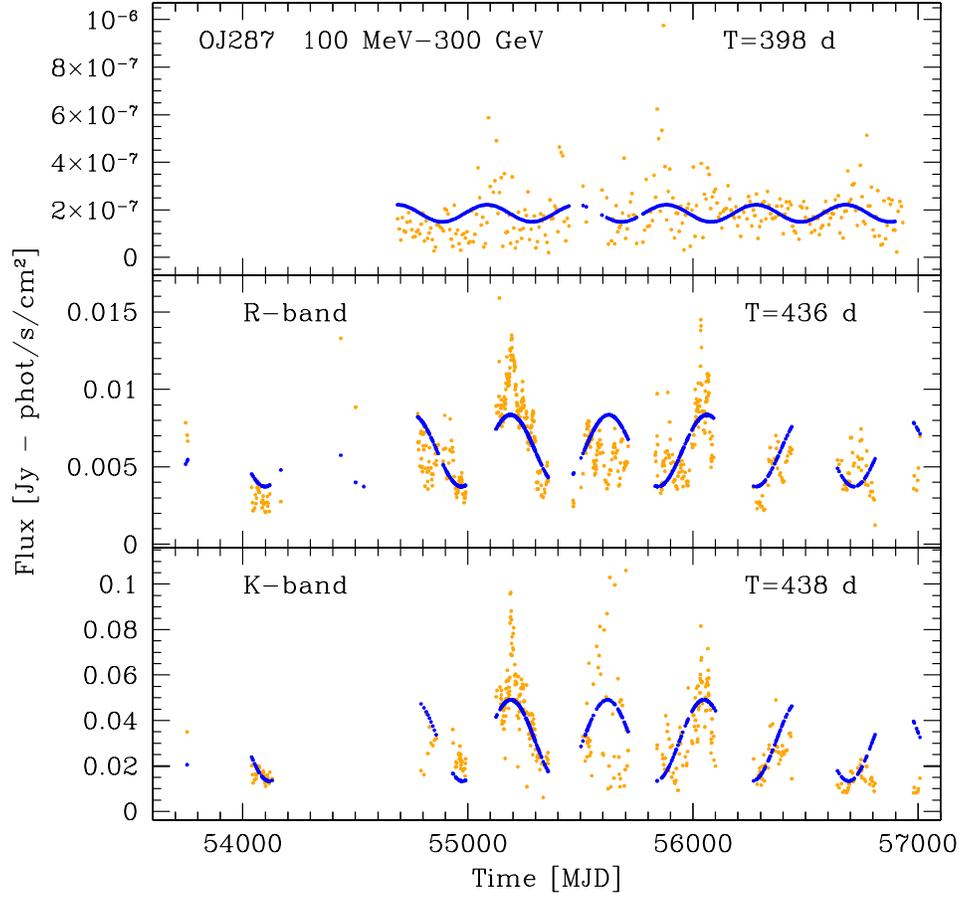}
	\caption{ \label{3lc_oj287}  
	 Same of Figure \ref{3lc_pks0537} for OJ 287.
	The amplitudes of the sinusoidal curves are 
	 A=3.60 $\cdot$ 10$^{-8}$  photons $\cdot$ s$^{-1}$ $\cdot$ cm$^{2}$ \ in 100 MeV-300 GeV (T=398 d), \ 
	 A=0.0023  Jy  in R (T=436 d) \ 
	 and A=0.0179 Jy in K band (T=438 d).			
	 }
\end{figure*}
\begin{figure*}
\centering 
\includegraphics[trim=0cm 1cm 0cm 0cm,clip,width=.8\textwidth]{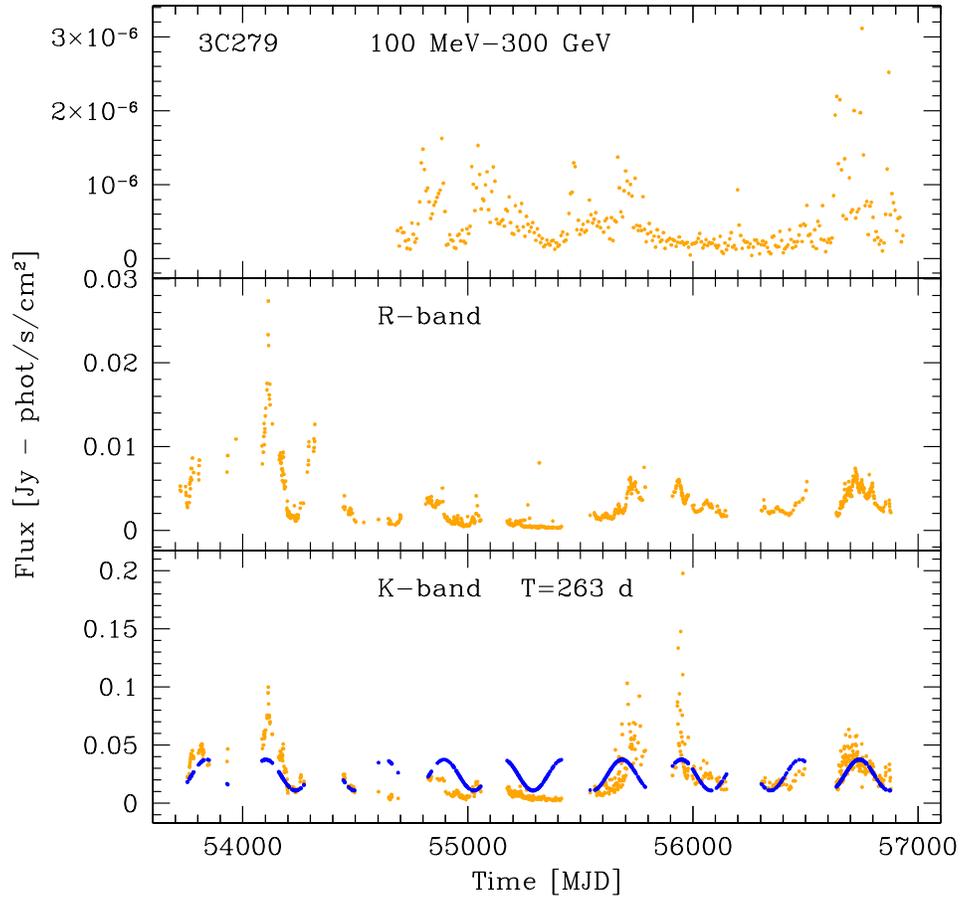}
	\caption{ \label{3lc_3c279}  
	 Same of Figure \ref{3lc_pks0537} for 3C 279. 
	 The amplitude $A$ for the sinusoidal curve with a 263 d period in K-band is 0.0133 Jy.	
	 }
\end{figure*}
\begin{figure*}
\centering 
\includegraphics[trim=0cm 1cm 0cm 0cm,clip,width=.8\textwidth]{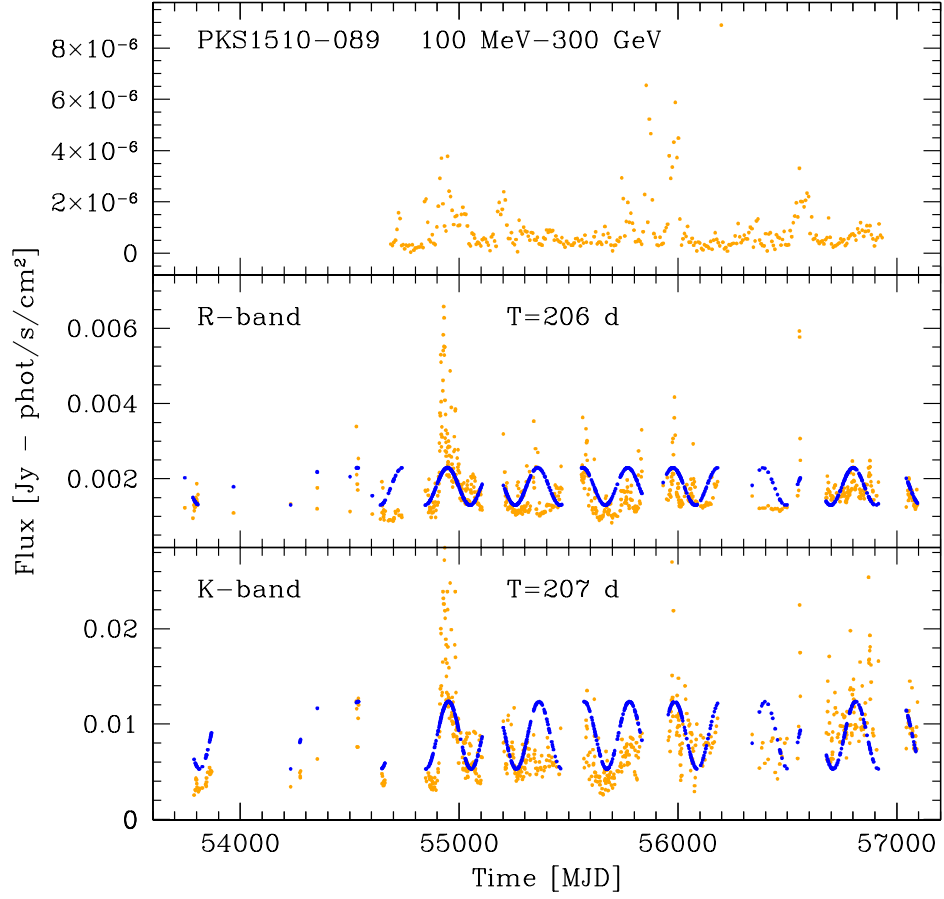}
	\caption{ \label{3lc_pks1510}  
	 Same of Figure \ref{3lc_pks0537} for PKS 1510-089. The prominent flare occurring 
	 on $\sim$ 54960 MJD is partially cut for an easier visualization of the data.
	  Amplitudes of the sinusoidal curves are 
	 A=0.000497  Jy  in R (T=206 d),  and A=0.0048 Jy in K band (T=207 d and T=474 d).			
	 }
\end{figure*}
\begin{figure*}
\centering 
\includegraphics[trim=0cm 1cm 0cm 0cm,clip,width=.8\textwidth]{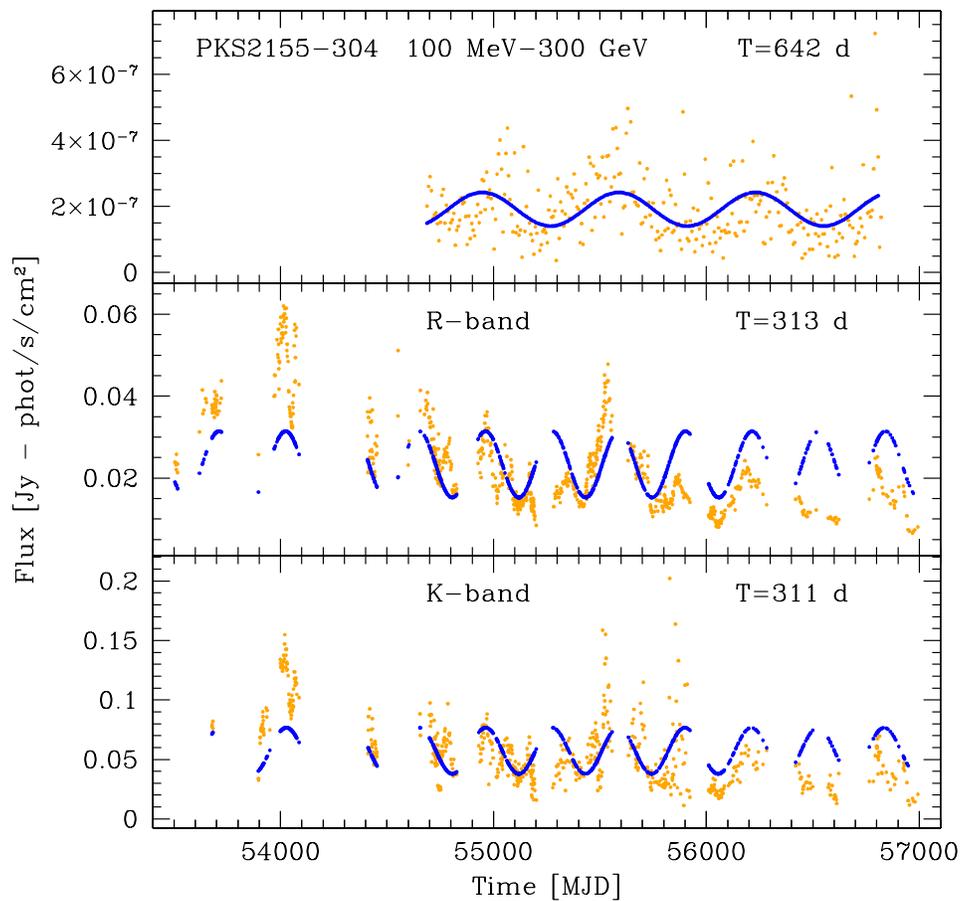}
	\caption{ \label{3lc_pks2155}  
	 Same of Figure \ref{3lc_pks0537} for PKS 2155-304. 	
	 Amplitudes of the sinusoidal curves are 
	 A=3.792 $\cdot$ 10$^{-7}$  photons $\cdot$ s$^{-1} \cdot$ cm$^{2}$ 
	  in 100 MeV-300 GeV (T=642 d), 
	A=0.00809  Jy  in R (T=313 d), 
	 and A=0.0194  Jy in K band (T=311 d).			
	 }
\end{figure*}

The starting point of our analysis are the light curves obtained combining 
REM  (2005-2012) and SMARTS photometry  (2008-2014) in V, R, J, K bands,  see Figures 
\ref{3lc_pks0537}-\ref{3lc_pks2155}
for the cases of R and K filters.
Note that contrary to our REM data,  SMARTS' photometry is taken at face value,
 as archived in the public web site.
 \textit{Fermi} data\footnote{\texttt{http://fermi.gsfc.nasa.gov/ssc/data/access/lat/msl\_lc/}}
  are also reported in Figure \ref{3lc_pks0537}-\ref{3lc_pks2155}
  in the 100 MeV - 300 GeV band.
The procedures for constructing gamma ray curves are complex but rather
 standardized\footnote{The data have been analyzed by using the standard 
  Fermi LAT ScienceTools software package, see\\
  \texttt{http://fermi.gsfc.nasa.gov/ssc/data/analysis/\\
  documentation/Cicerone/}}, 
  and  fully described e.g. in \cite{Abdo2010}. Note that these curves are 
 not corrected for background.

 The search for  (quasi) periodicities in the light curves of active galactic nuclei
  is notoriously an arduous problem, as it was pointed out in the seminal paper 
  by \cite{Press1978}, who considered the X-ray light curve of 3C 273 
   and indicated a number of caveats, which should be taken into account,
  before assessing  the reality of a periodicity.  
  In our case four main points should be examined:
  \begin{itemize}
  
\item Our optical sampling is rather irregular, as usual in ground based observations.
 On the other hand $\gamma$-ray light curves are essentially evenly spaced. 
 \item The light curves are affected by frequency dependent  \textit{red} noise.
   \item The total duration $T_{tot}$ of the monitoring is $\leqslant$ 9 yr in the optical, 
  $\leqslant$ 6 yr in gamma rays, constraining the minimum frequency 
that can be searched for.

\item Flares or periods of high activity can affect the analysis, requiring a careful check of the results.
 \end{itemize}

   The problem of the presence of red noise in evaluating a periodogram 
  has been discussed in detail in the case of X-ray light curves,  mainly in the context
   of galactic sources.    
The case of time series not evenly spaced typically requires the use of interpolation techniques.
Their application implies that the interpolated data points are no longer independent
and may introduce a significant additional bias, which leads to an enhancement of low-frequency 
components at the expense of higher frequency ones
\citep[e.g.][]{Schulz1997}.

The assessment  of  the reality of periodicities cannot be  carried out without firm
procedures  to measure a significance  against the background
noise.
    In particular the problem of the red noise modeling  was examined
   by,  i.e., \cite{Israel1996}, \cite{Vaughan2005,Vaughan2010}, 
\cite{vanderKlis1989a,vanderKlis1989b}, \cite{Zhou2002}, \cite{Timmer1995},
although usually for the simpler case of evenly sampled time-series.

A procedure for the study of periodicities in non-equally spaced light curves and 
    affected by red noise was developed by \cite{Schulz2002}
   with reference to paleoclimatic time 
   series\footnote{Details on the relevant software "REDFIT" can be found at
\texttt{http://www.geo.uni-bremen.de/geomod/staff/mschulz/\\\#software2}}. 
In this procedure  a  first-order
autoregressive (AR1) process is applied to model the red-noise 
\citep{Hasselmann1976}. 
This avoids interpolation in the time domain with the introduced bias.
The AR1 technique can give a good description of rather smooth and 
regular time-series, as it is the case of the Fermi light-curves we are here
 considering. Our optical light-curves are, on the contrary, much more difficult 
 to model due to periods of high-activity and/or flares introducing order of 
 magnitudes variations together with a highly irregular sampling. 
 Nevertheless, the results of these analyses can provide useful hints about 
 the significance of a detected possible periodicity, in particular when these 
 periods are also present in the Fermi data.

\begin{figure*}
\centering 
\includegraphics[trim=0cm 0cm 0cm 0.5cm,clip,width=0.45\textwidth]{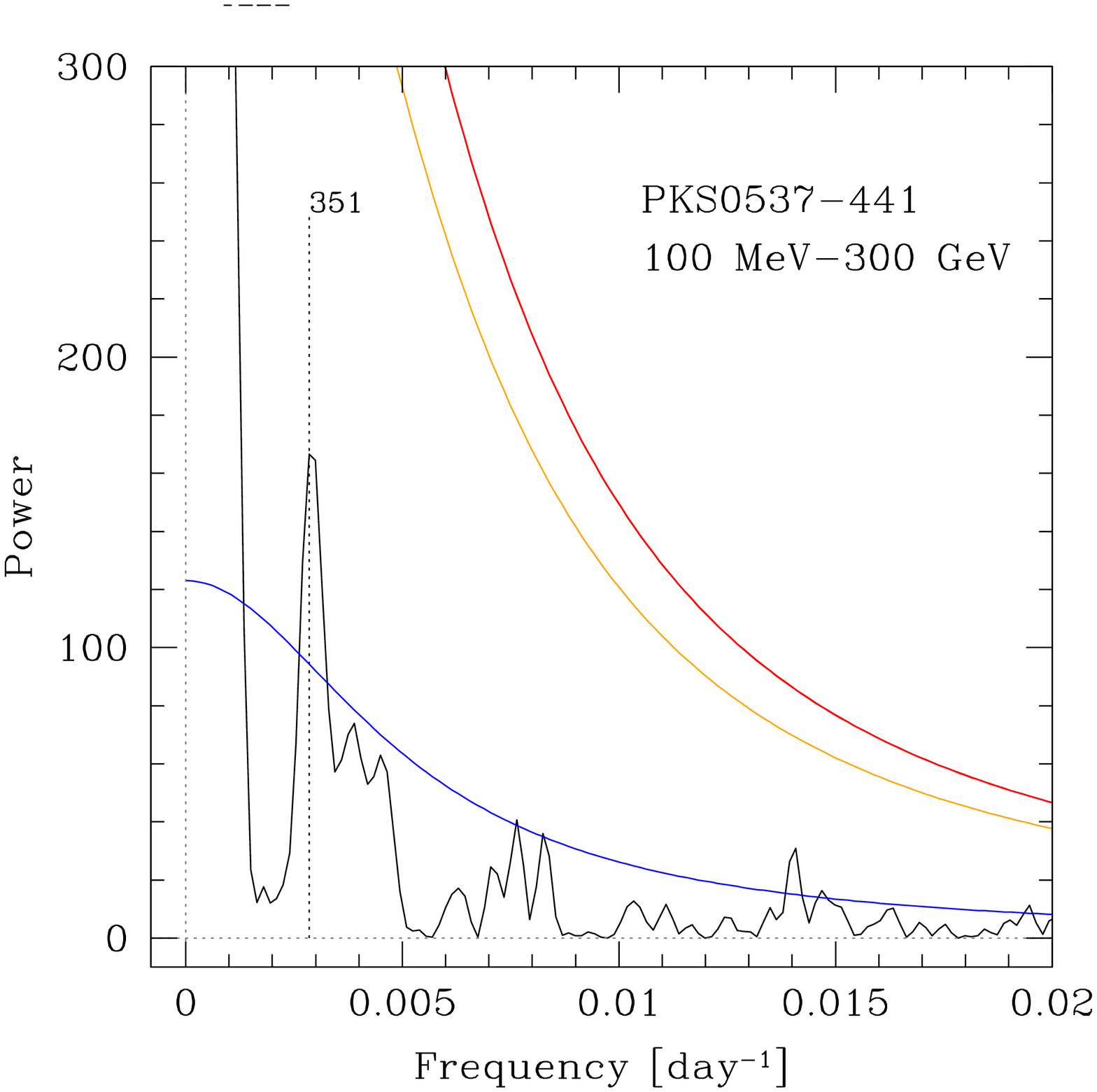}
\includegraphics[width=.45\textwidth]{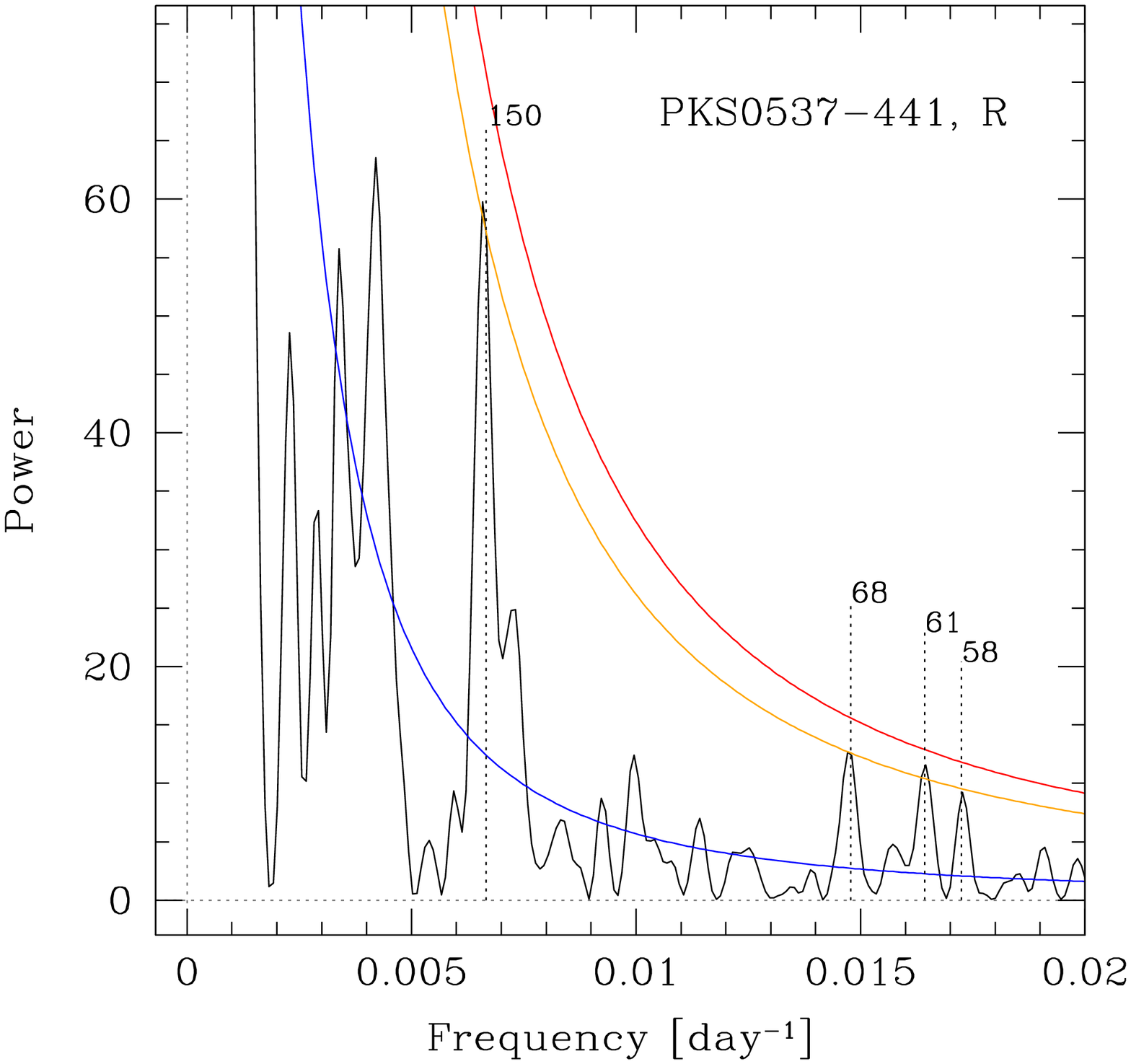}
\includegraphics[width=.45\textwidth]{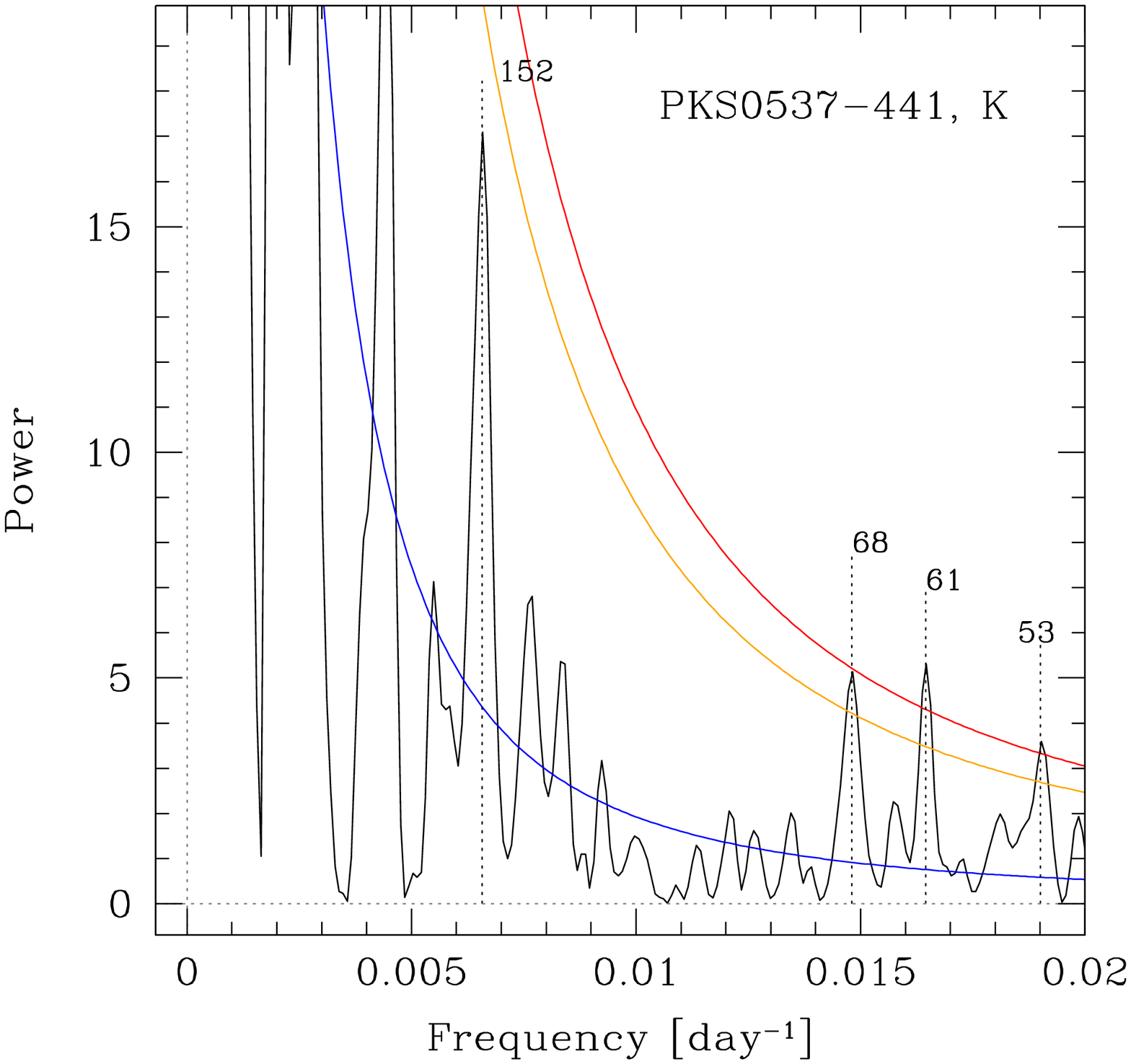}
	\caption{ \label{pow_pks0537}  
		Bias-corrected power spectra  (black line) of the blazar sample  
		in 100 MeV-300 GeV from the Fermi
		$\gamma$-ray light curves, and 	
		in K and R bands   from REM + SMARTS photometry
		(see also Table \ref{top_hits_pks0537}-\ref{top_hits_pks2155}).
		The power is the output obtained by the procedure  of \citet[REDFIT]{Schulz2002}
		normalized with the variance.
		 Curves in each panel, starting from the bottom, are: the theoretical red-noise spectrum,
		  the 99.0\%  (2.5$\sigma$) and 99.7\% (3$\sigma$) $\chi^{2}$ 
		 significance levels. 
               	Periods in days corresponding  to the prominent peaks are marked.
				}
		\end{figure*}

\begin{figure*}
\centering 
\includegraphics[width=.45\textwidth]{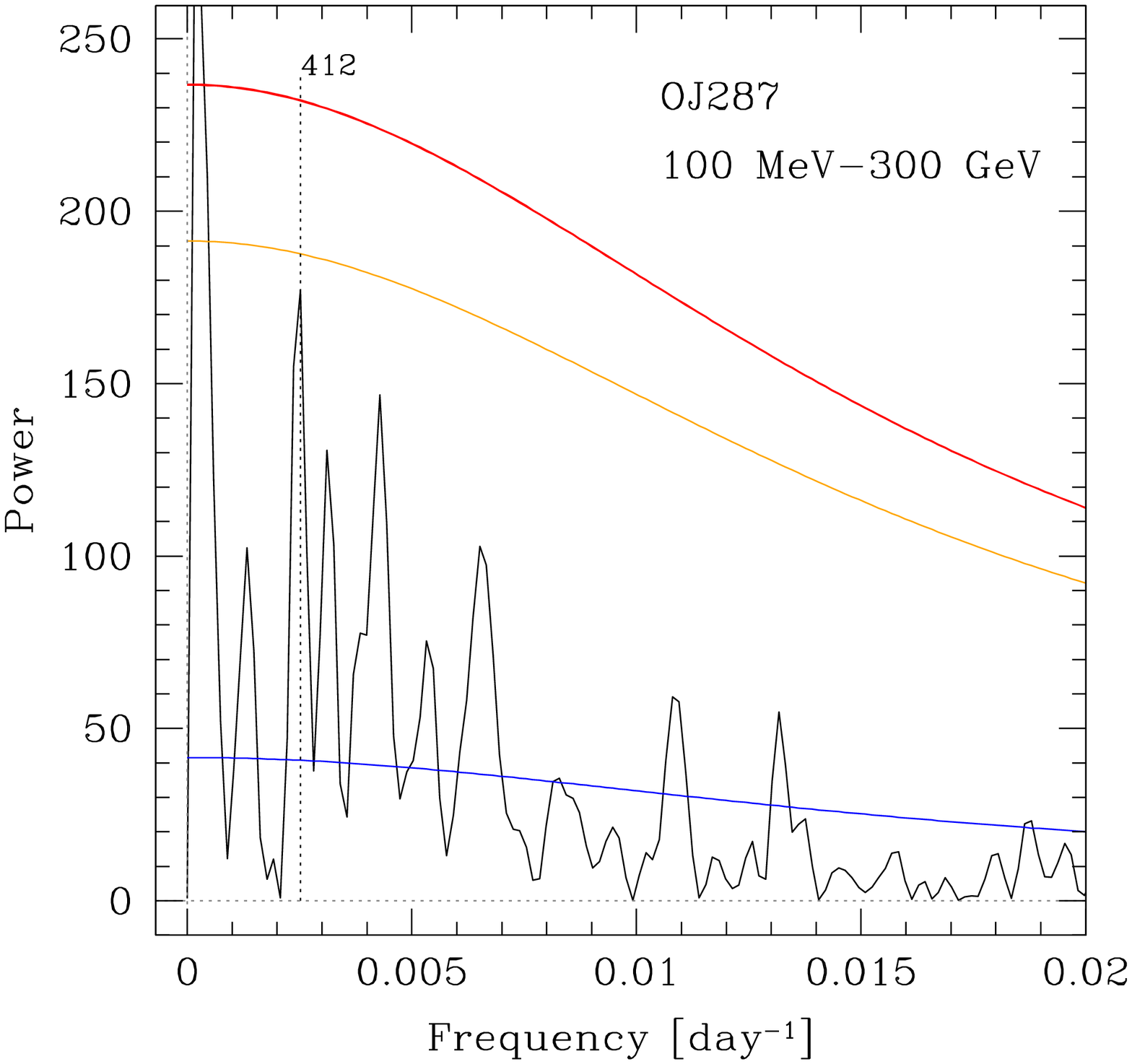}
\includegraphics[width=.45\textwidth]{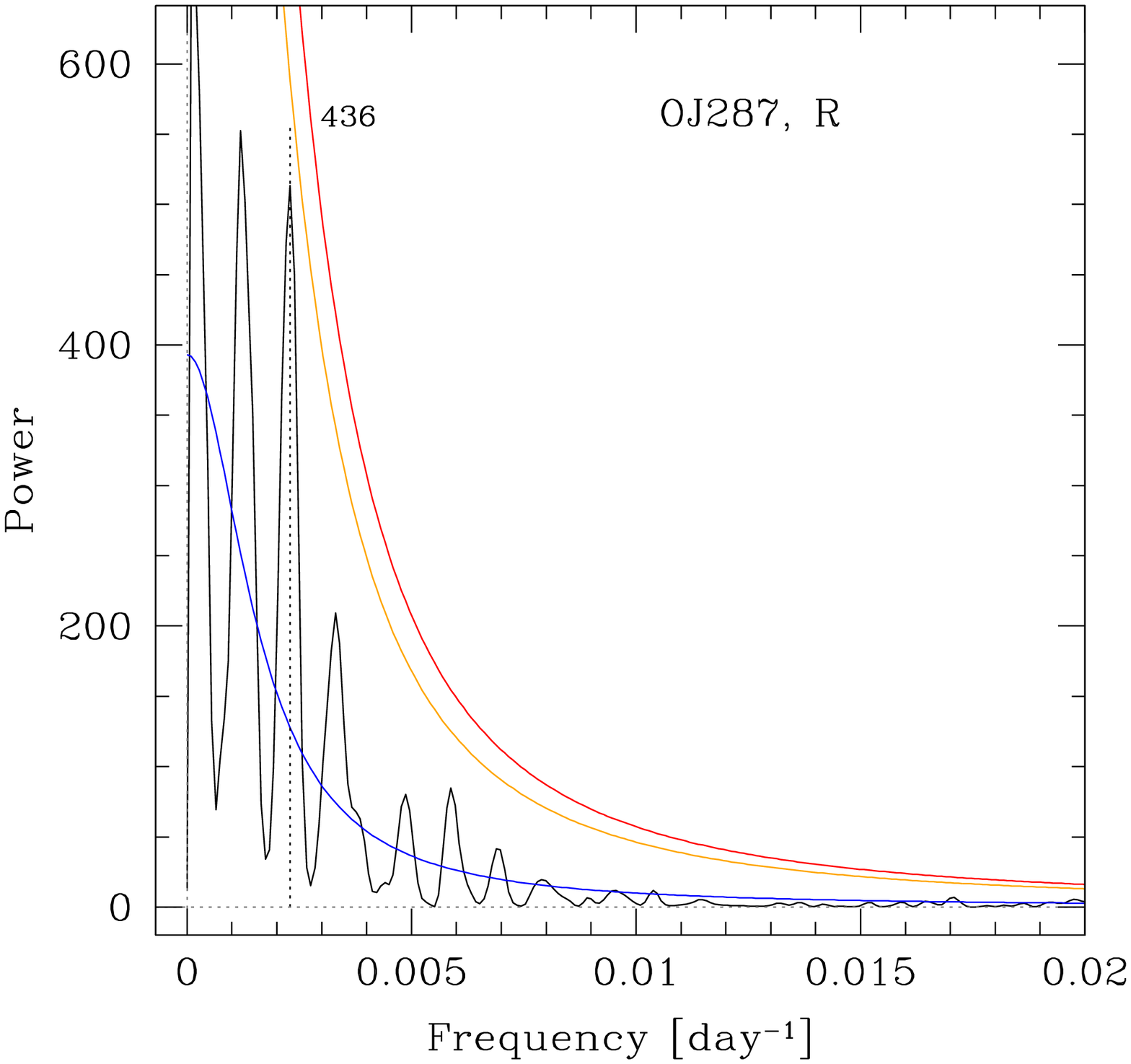}
\includegraphics[width=.45\textwidth]{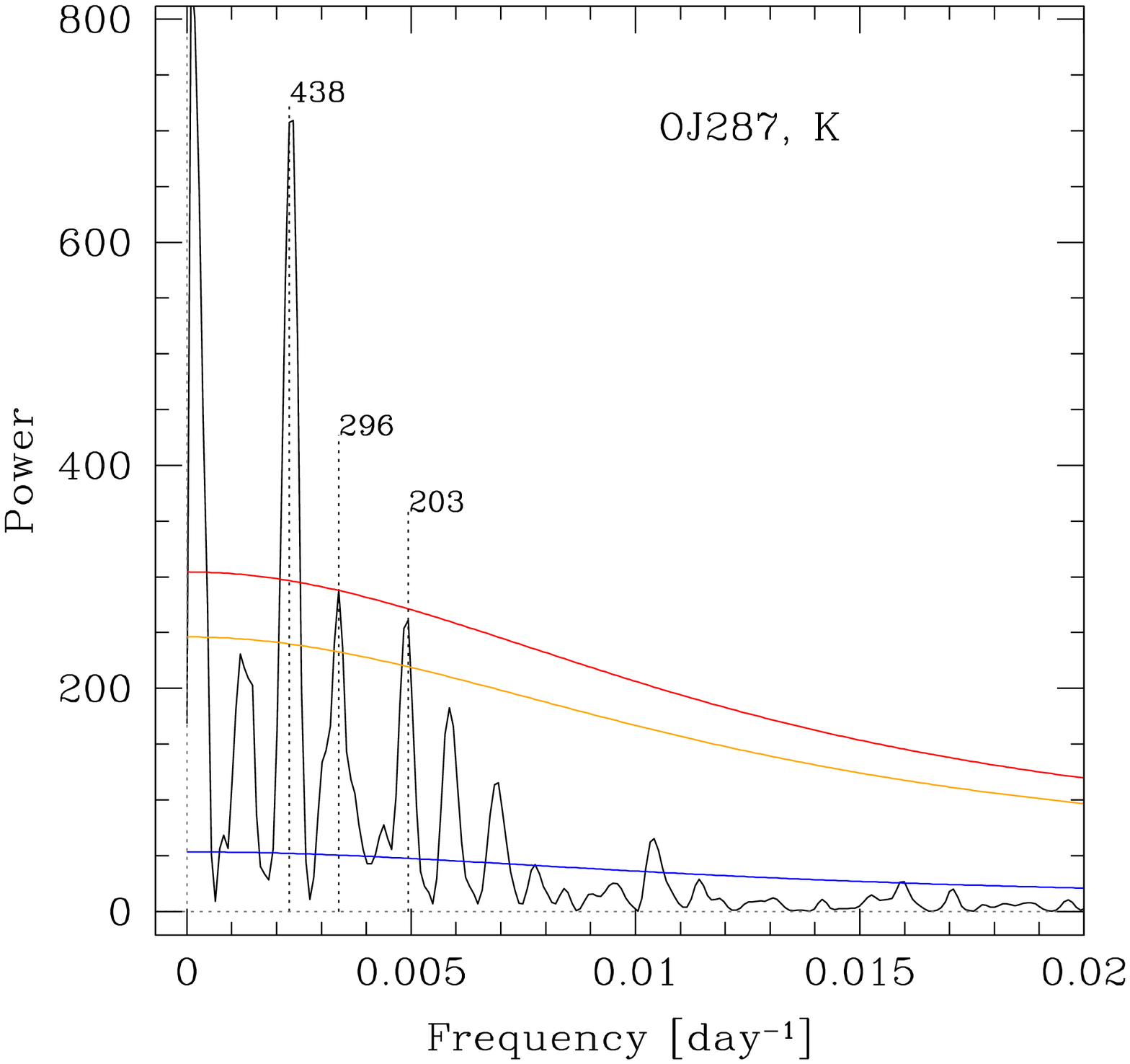}
	\caption{\label{pow_oj287} The same of Figure \ref{pow_pks0537} for OJ 287.}
\end{figure*}

\begin{figure*}
\centering 
\includegraphics[width=.45\textwidth]{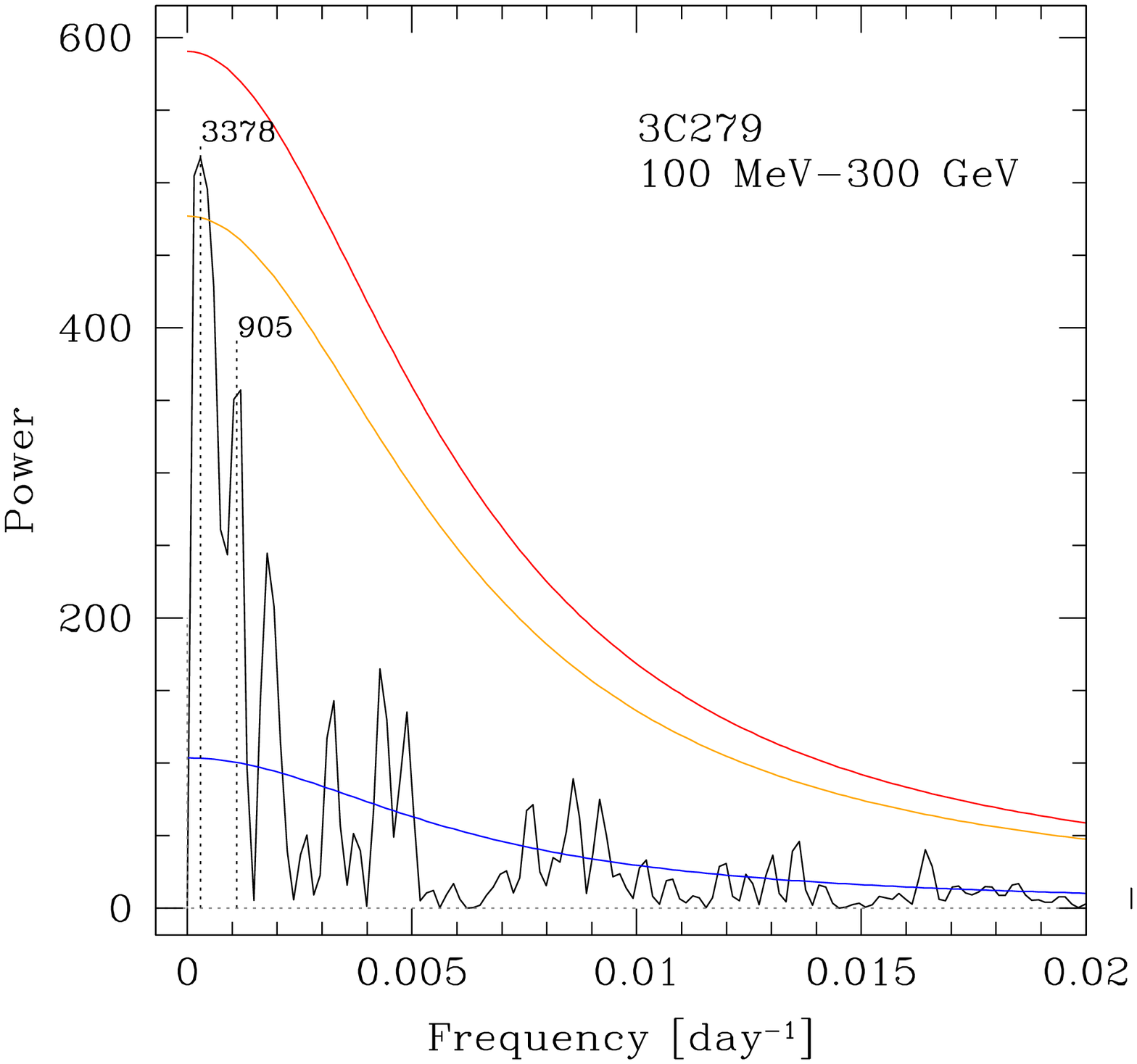}  
\includegraphics[width=.45\textwidth]{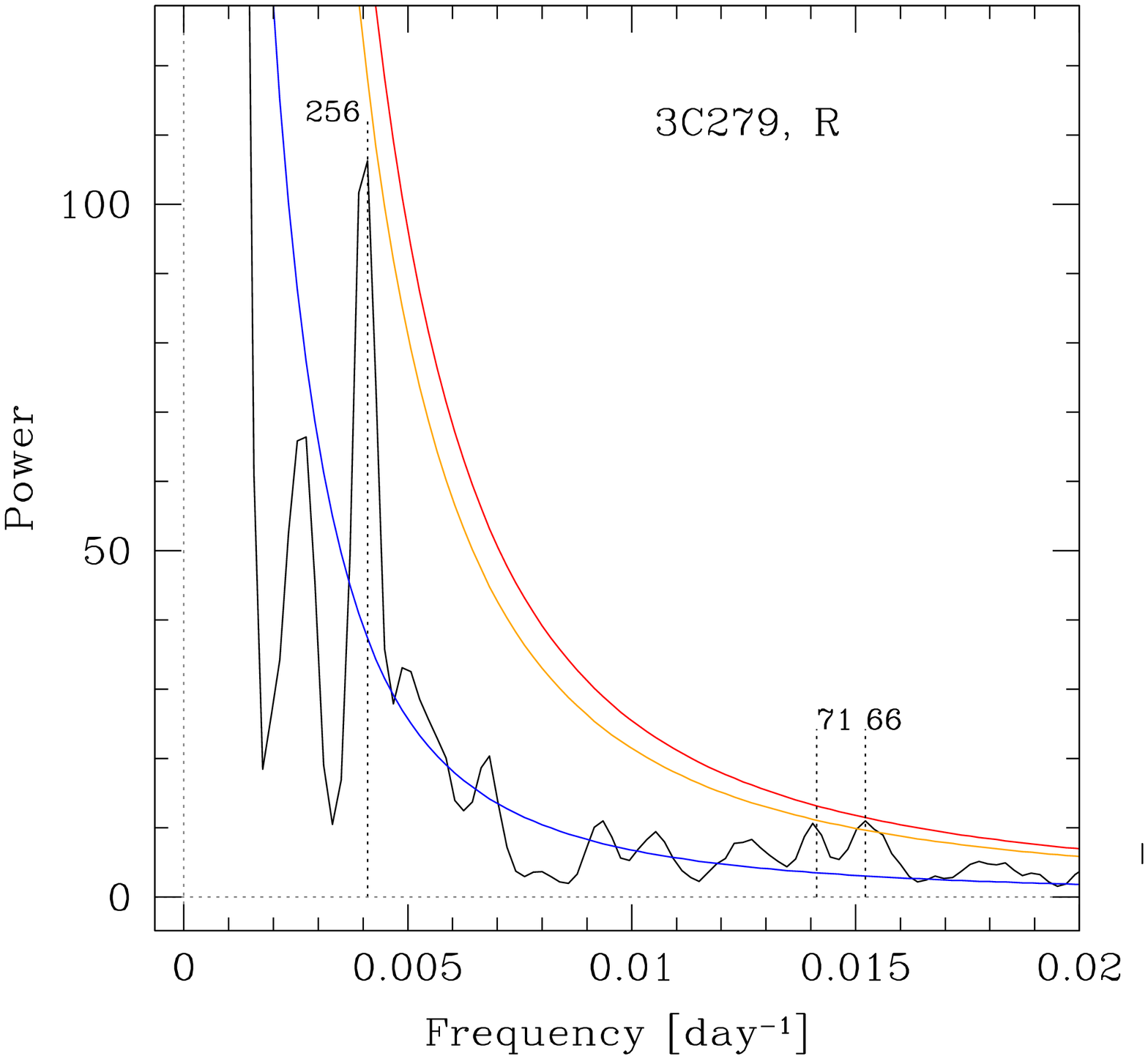}
\includegraphics[width=.45\textwidth]{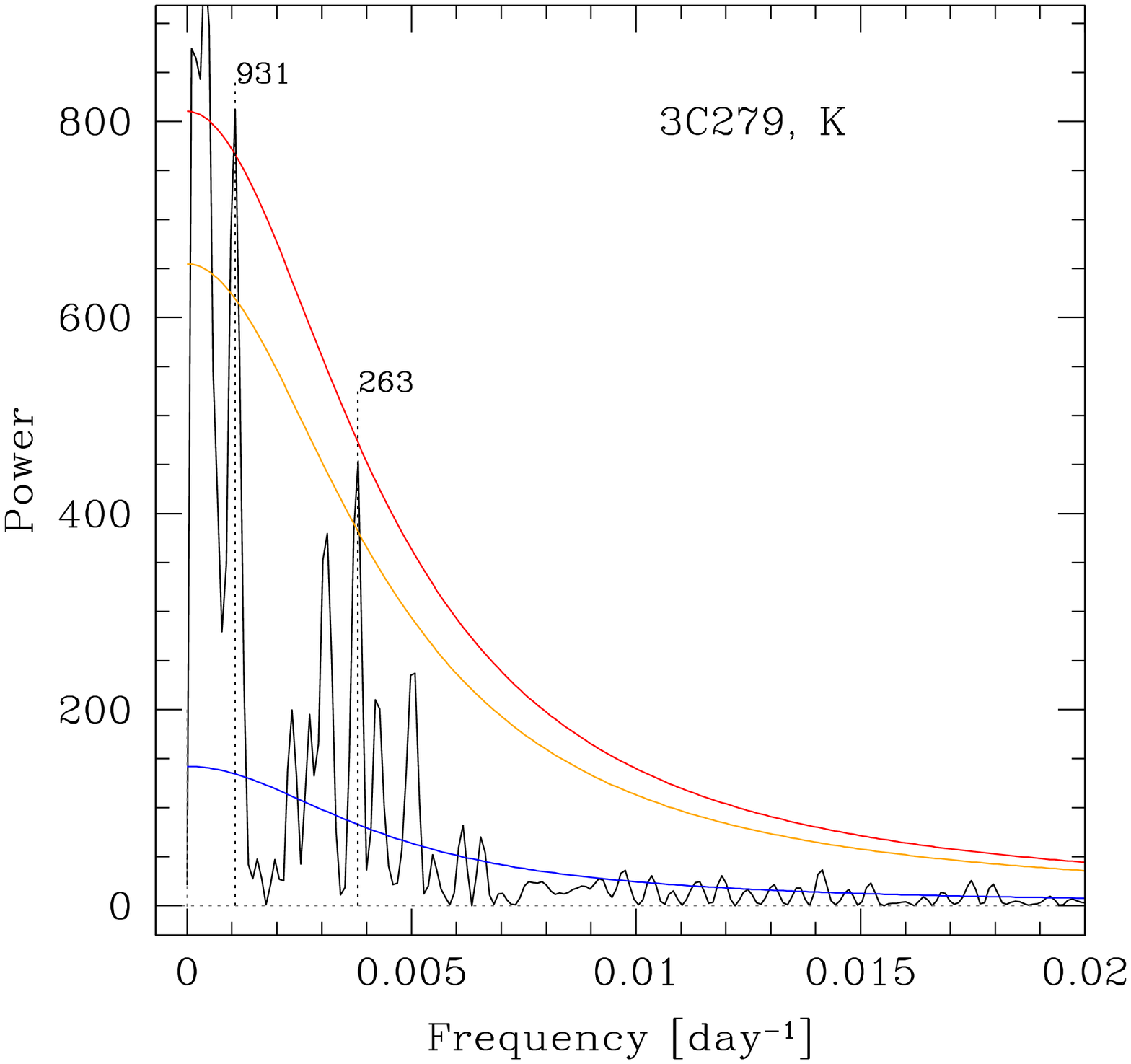}
	 \caption{\label{pow_3c279} The same of Figure \ref{pow_pks0537} for 3C 279.}
 \end{figure*}

\begin{figure*}
\centering 
\includegraphics[width=.45\textwidth]{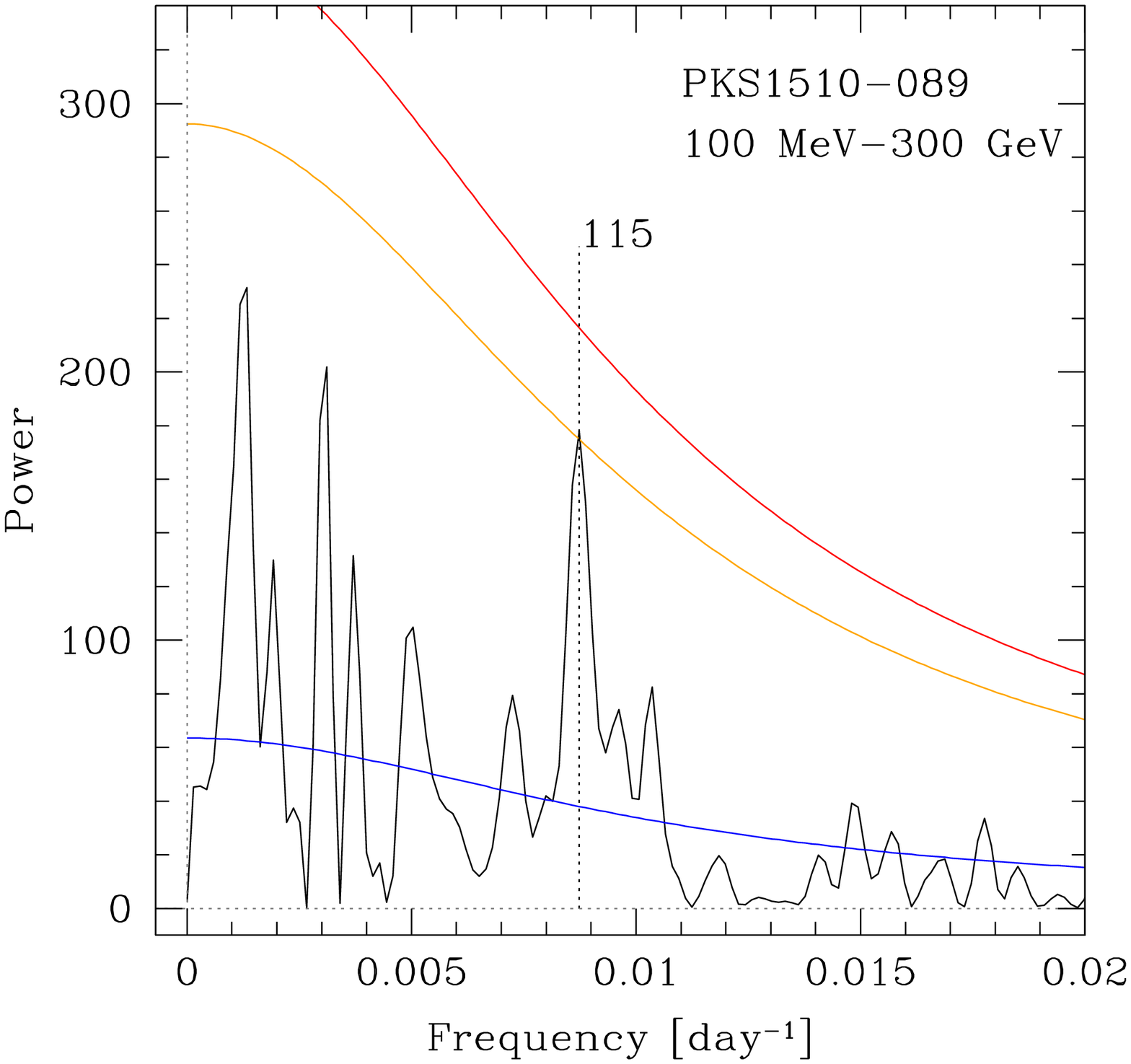}
\includegraphics[width=.45\textwidth]{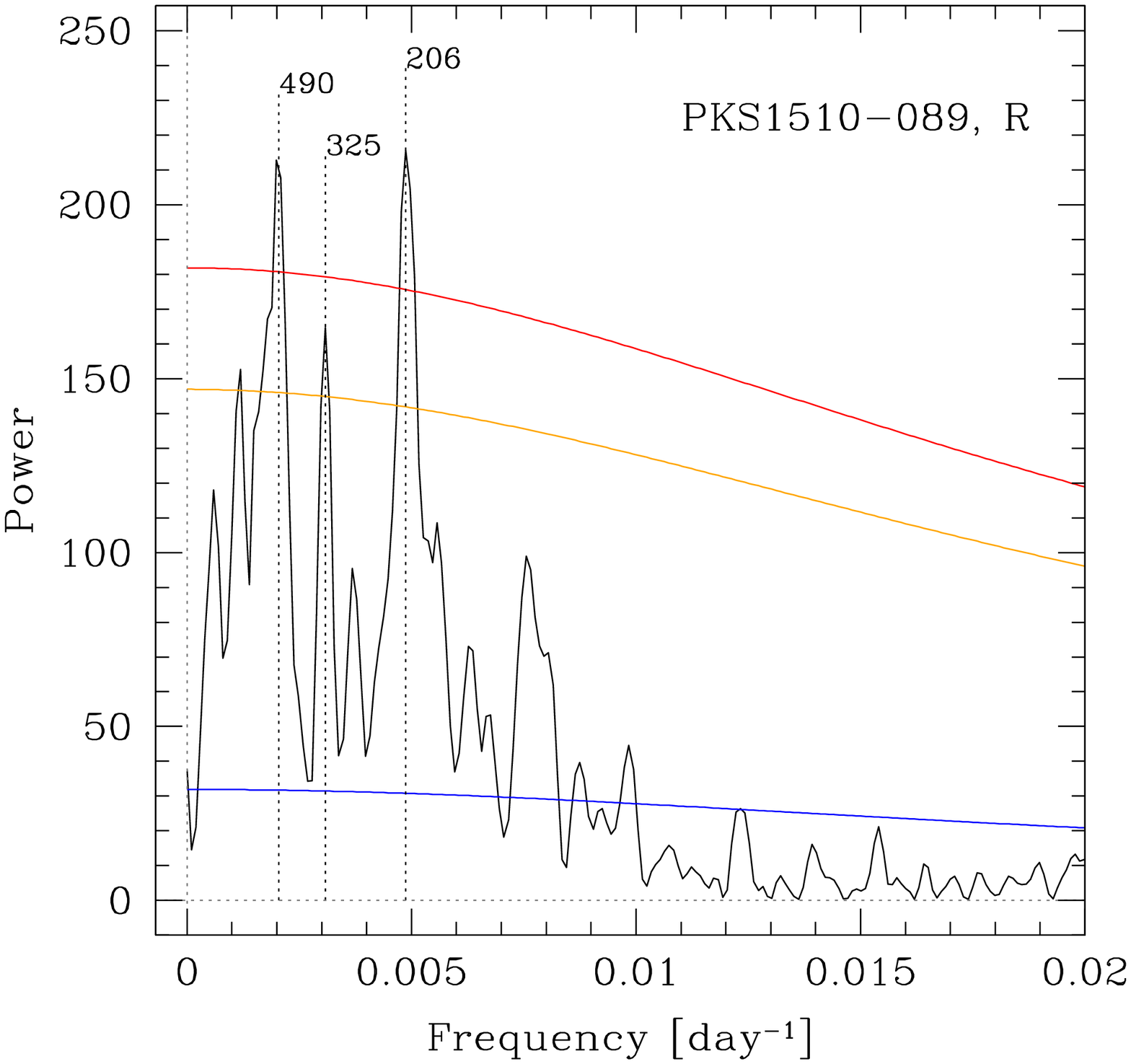}
\includegraphics[width=.45\textwidth]{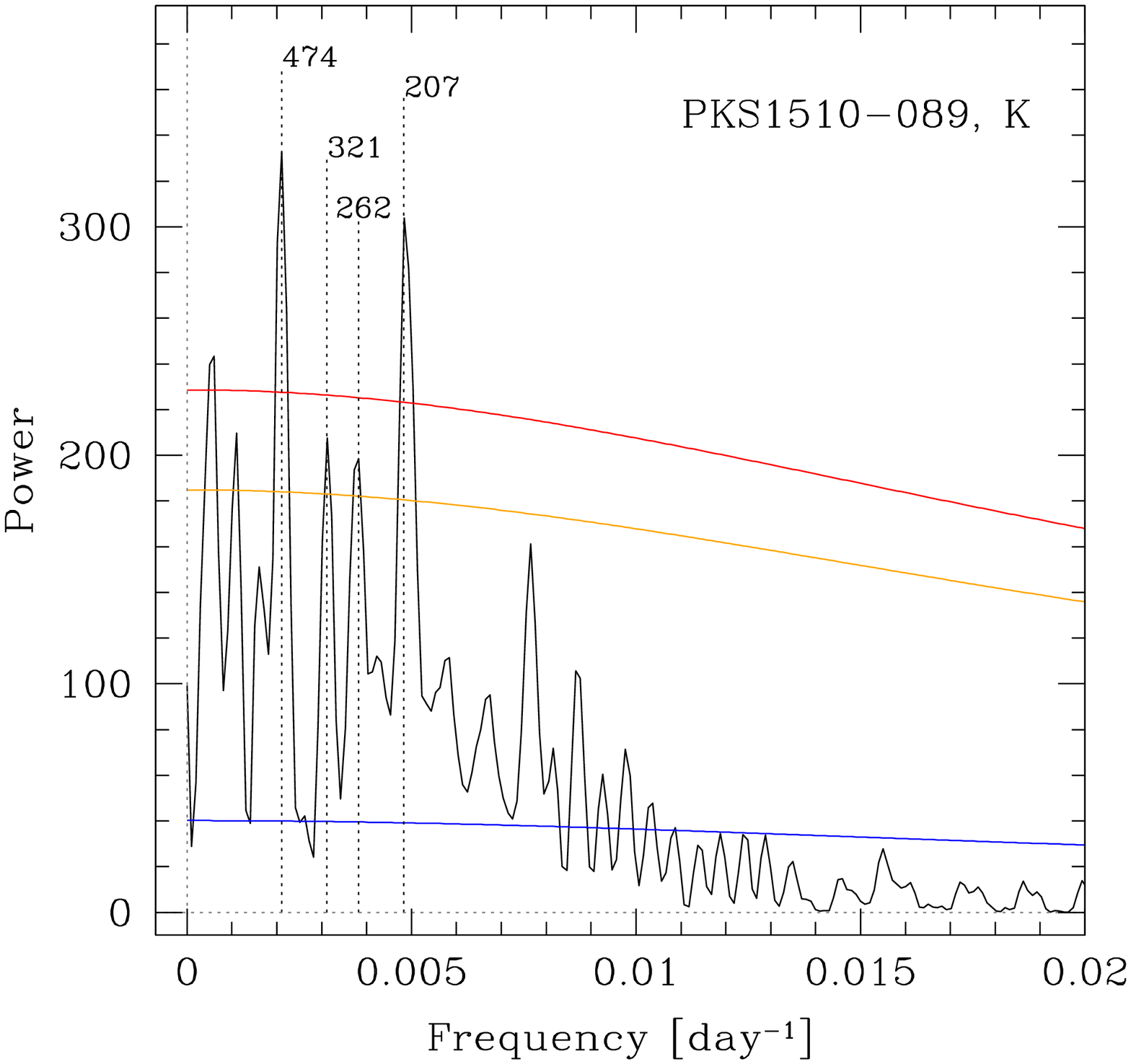}
           \caption{\label{pow_pks1510} The same of Figure \ref{pow_pks0537} for PKS 1510-089.}
\end{figure*}

\begin{figure*}
\centering 
\includegraphics[width=.45\textwidth]{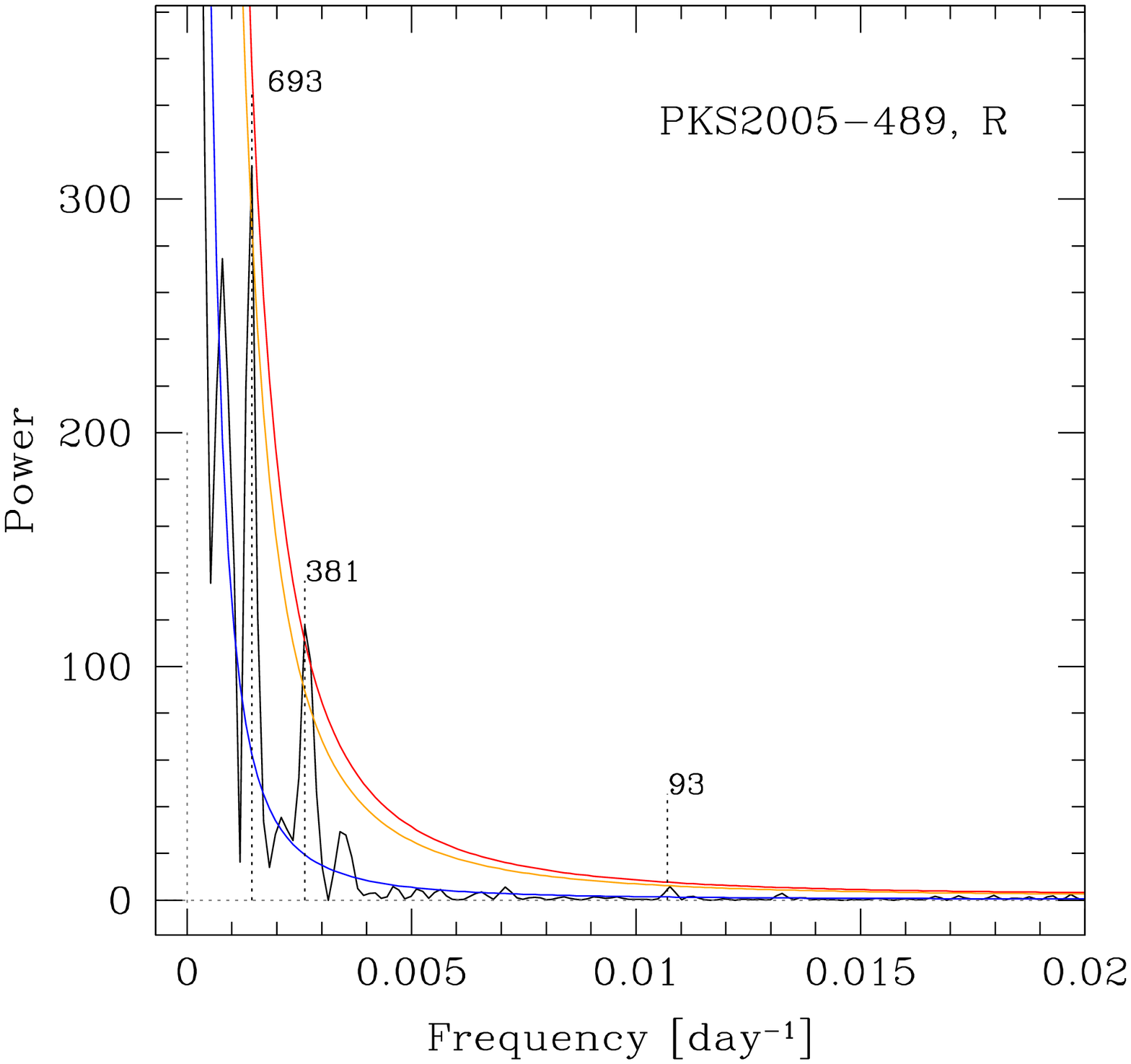}
\includegraphics[width=.45\textwidth]{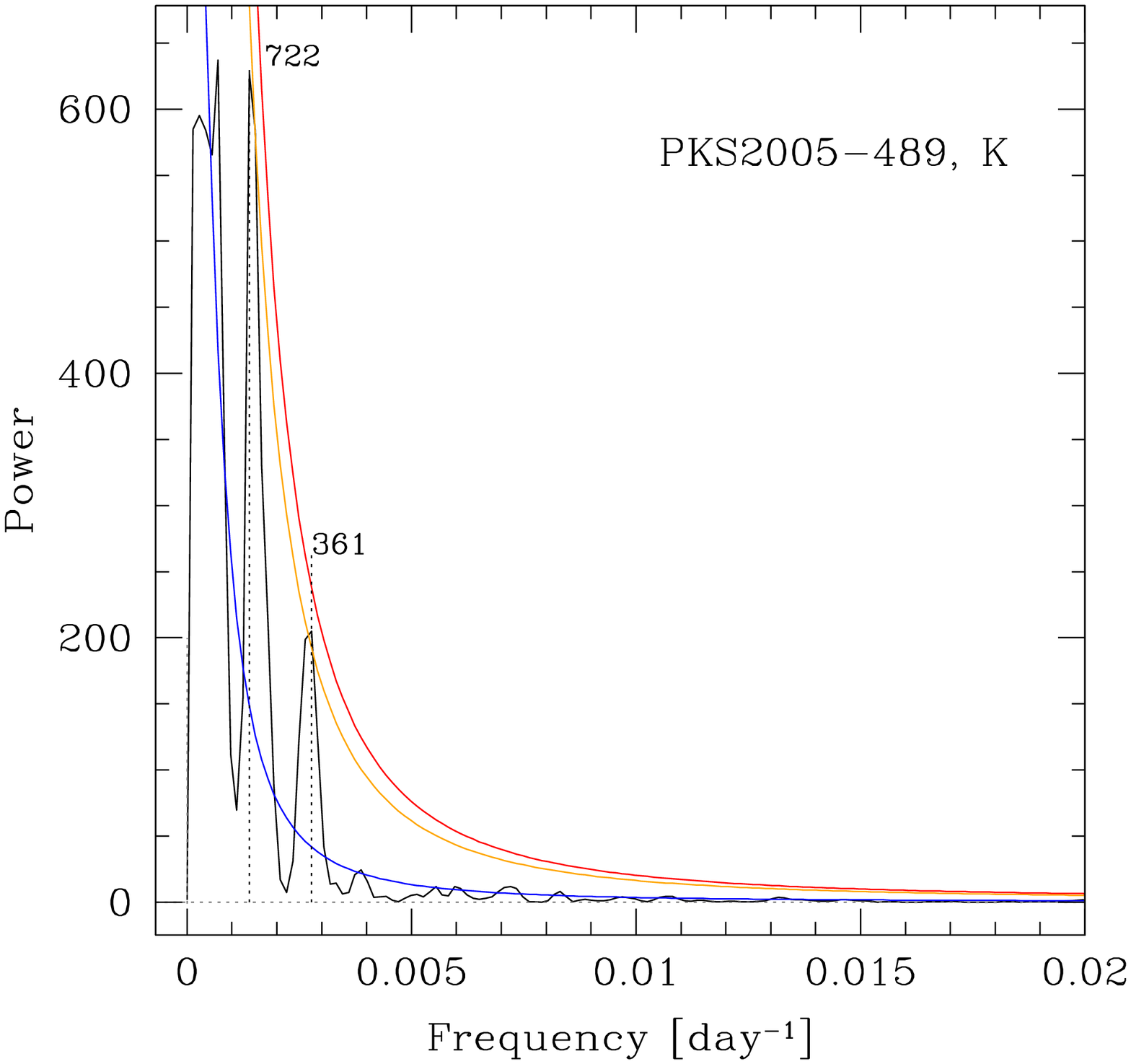}
	 \caption{\label{pow_pks2005} The same of Figure \ref{pow_pks0537} for PKS 2005-489.
	 }
\end{figure*}

\begin{figure*}
\centering 
\includegraphics[width=.45\textwidth]{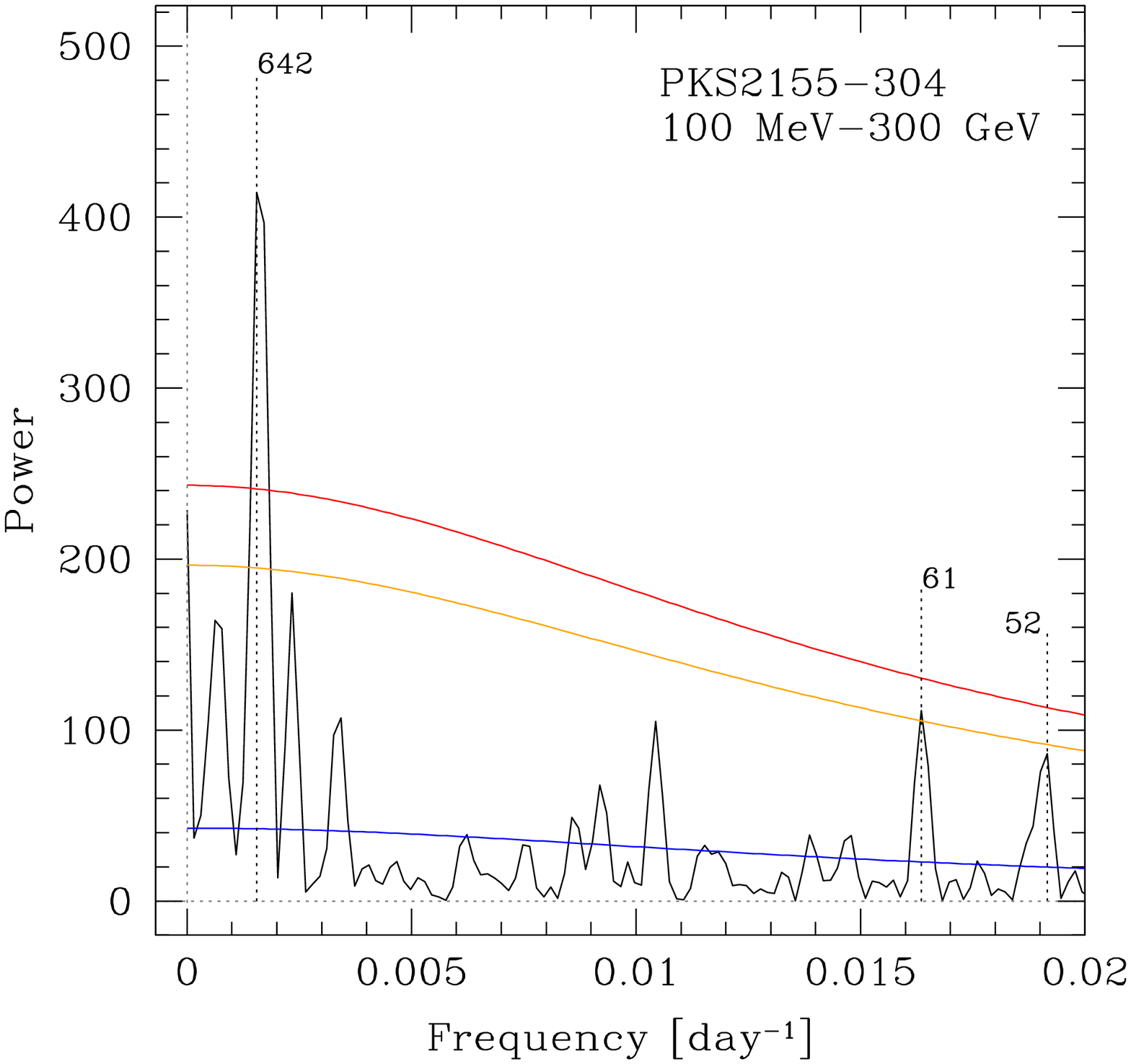}
\includegraphics[width=.45\textwidth]{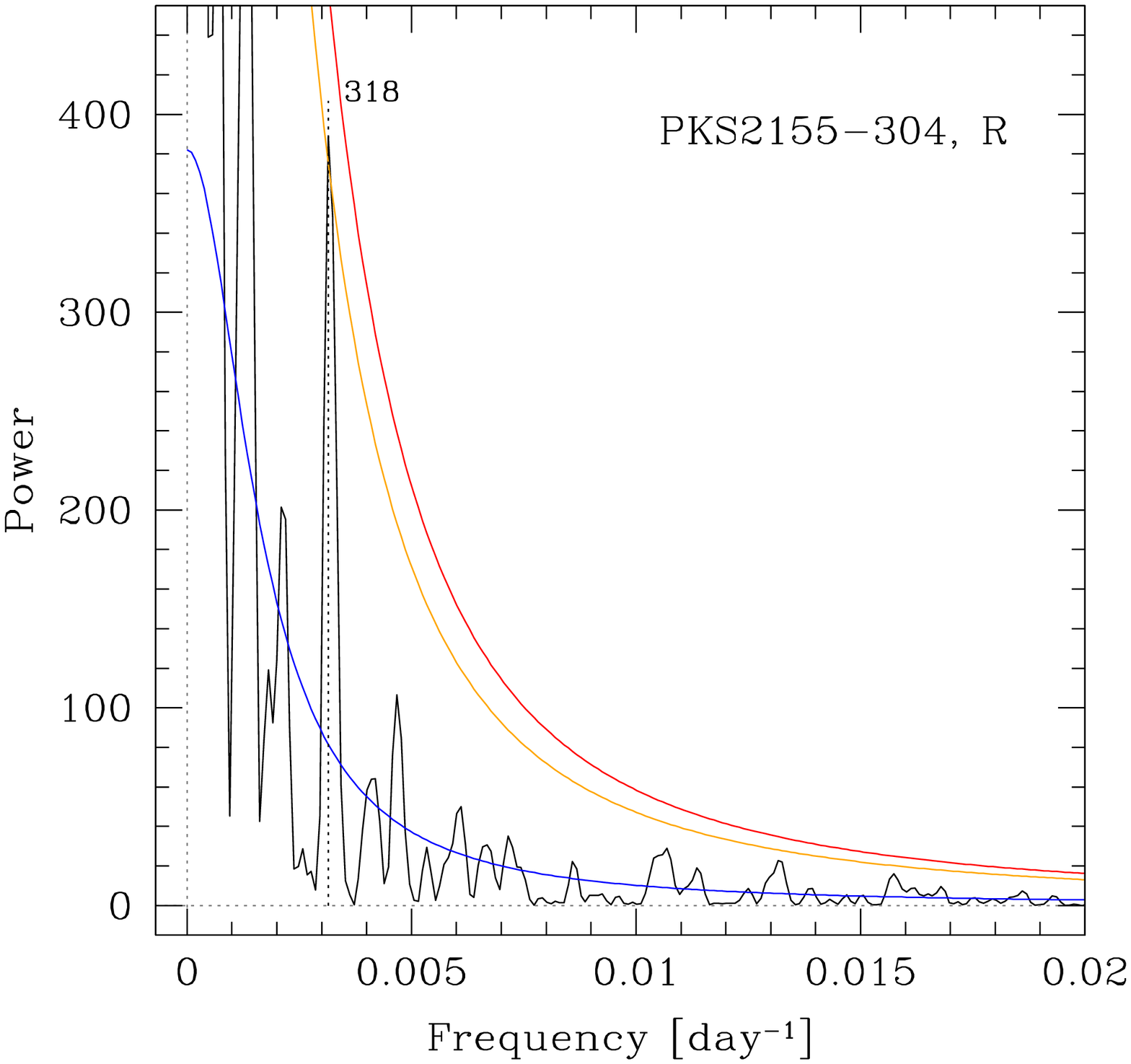}
\includegraphics[width=.45\textwidth]{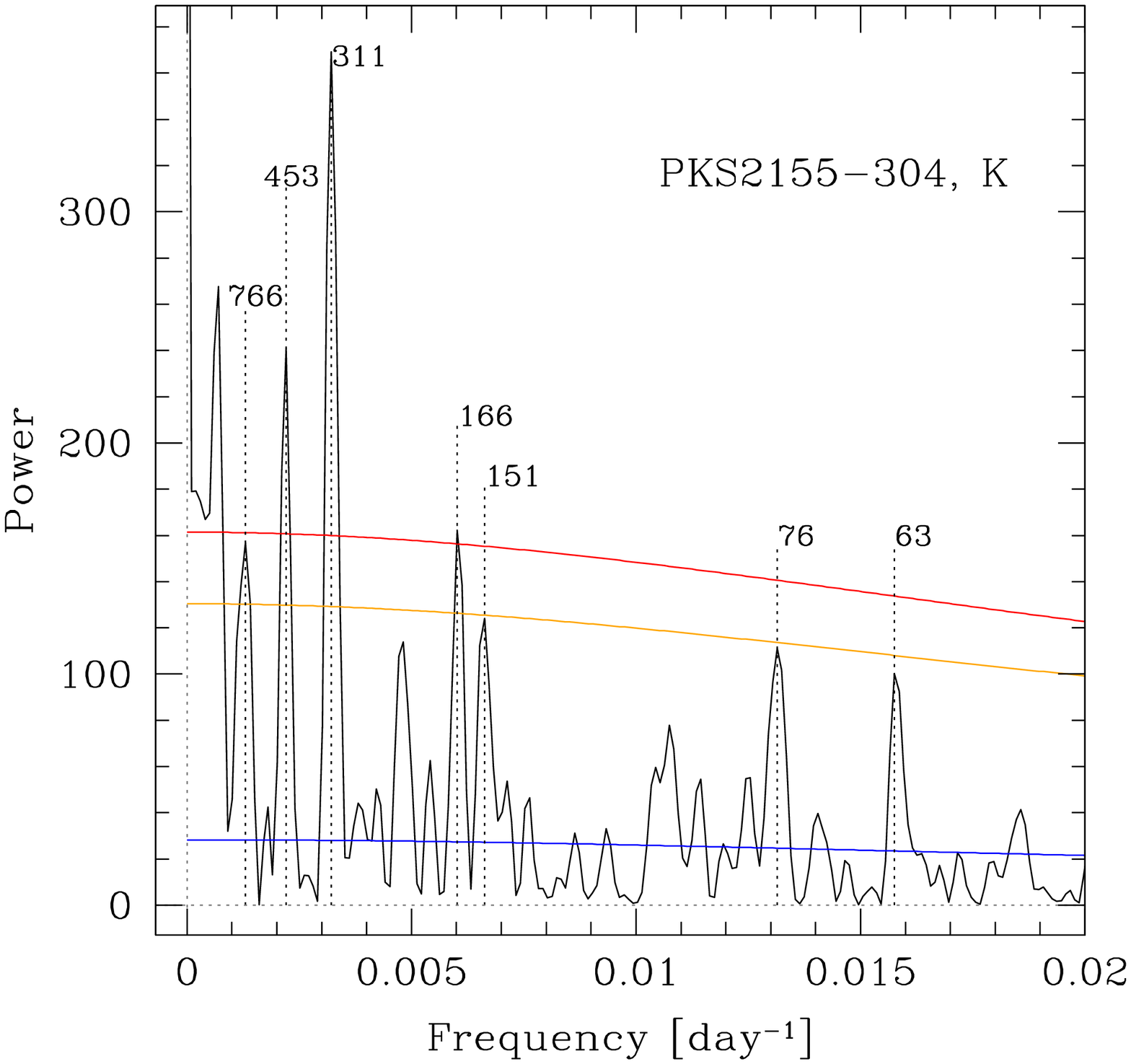}
	 \caption{\label{pow_pks2155} The same of Figure \ref{pow_pks0537} for PKS 2155-304.}
 \end{figure*}

We started with light curves with binning of 1 day in the optical bands, 
  and 1 week in the $\gamma$-rays (see Figure \ref{3lc_pks0537}-\ref{3lc_pks2155}).
We considered $\gamma$-rays bins with test statistics \citep{Mattox1996}
 $TS>$ 4, corresponding to a $\sim$ 2$\sigma$ detection.
Note that  for the large majority of cases the detections are much more significant than this limit. 
We have chosen a maximum frequency for the period analysis 
  corresponding to $\sim$ 20 days.  
  In fact here we are not interested in the short time scale variability.
   The procedure yields (see Figures \ref{pow_pks0537}-\ref{pow_pks2155}):
   \begin{itemize}
   \item  Lomb and Scargle periodograms \citep{Scargle1982},
   which account for the unevenly spaced photometry.
   \item Modeling  of  red noise continuum.
   \item  Bias-corrected spectra and  significance ($S$) level curves: we have chosen
    $S=99.0\%$ (2.5$\sigma$)  and $S= 99.7\%$ (3$\sigma$). 
       \end{itemize}
       
The significance is given by the comparison of the periodogram with that based on the
auto-regressive model (see above)  to test the null hypothesis 
that the observed time series can be fully accounted by the noise.

In Tables \ref{top_hits_pks0537}-\ref{top_hits_pks2155} we report the \textit{observed} periods 
corresponding to peaks with $\gtrsim$ 99\% significance.  
We evaluated the period uncertainty following  \cite{Schwar1991},
searching for the Mean Noise Power Level (MNPL)  in the vicinity of the
investigated  period  T.
The   1$\sigma$ confidence interval on T is the width 
of the peak in the power spectrum at  $p-$MNPL, where $p$ is the height
of the peak. 
The sinusoidal artificial light-curves with the most significant periods calculated using 
Vstar package\footnote{\texttt{http://www.aavso.org/vstar-overview}} are reported
in Figures \ref{3lc_pks0537}-\ref{3lc_pks2155}.
We checked for aliases derived from the interval sampling between observations or  
the sampling rate,
 which cause false peaks in the time analysis \citep[see e.g.][]{Deeming1975}. 
 We adopted the procedure in the Period Analysis 
 Software\footnote{\texttt{http://www.peranso.com}}  
and  found in
all the NIR-optical curves  evidences 
of  an alias period of 
 $\sim$ 370 d, which is representative of the year length. 
 The corresponding peak is negligible with respect to the other periodicities
 in all the sources with the exception of PKS 2005-489.
  Applying the same procedure to $\gamma$-ray
 light curves, no aliases have been  detected, as expected,
 except for a significant one at 7 d, denoting the sampling rate of the \textit{Fermi} data.
 The peaks, which appear marginally revealed in the spectral analysis for 
  alias detection, are marked in Tables \ref{top_hits_pks0537} - \ref{top_hits_pks2155}.  
  
We now examine the sources,  separately, discussing the most important possible
 periodicities detected by the analysis. As a general point, several periods
  appear to have a formal significance close or better than 99\%. 
  However considering the many possible sources of uncertainty 
  (mainly in the optical/NIR) it may be not sufficient  to provide a fully solid
   statement about their reliability. 

\textit{PKS 0537-441} --
In the R, J, K  bands there are peaks at $T_1 \sim$ 150 d with a significance $S\gtrsim$ 98\%, 
as reported in Table \ref{top_hits_pks0537},
and in the $\gamma$-rays a  peak appears at  351 d 
with $S$$\sim$ 90\%,
see Figure \ref{pow_pks0537}.
The most prominent peak  
at 1668 d	
 is close to the total observing time $T_{tot}$ and it is non reported in Table \ref{top_hits_pks0537}
 and Figure \ref{pow_pks0537}.
  
\textit{OJ 287} -- 
In the NIR-optical bands  plotted in Figure \ref{3lc_oj287} there are peaks 
 at $T_1 \sim $435 d ($S\gtrsim$ 99.7\% in V, J and K, see Table \ref{top_hits_oj}), which
within the errors may be related with  the $\gamma$-ray periodicity at $\sim$ 410 d ($S\sim$ 99\%).

\textit{3C 279} --
As it is apparent from  
Figure \ref{3lc_3c279},   there are 
significant peaks at 
$T_{1}  \sim 910$ d, which however is comparable with $T_{tot}$. $T_{1}$ 
is the most prominent one, which can be taken into account
 at low frequencies in the $\gamma$-ray spectrum ($S\gtrsim$ 95 \%).
 It is noticeable that peaks at $T_{1}$ and $T_{7} \sim$ 24 d are present both  in the periodograms
of the $\gamma$-ray data  and of the  NIR-optical bands
(in K-band   for $T_{1}$  $S$ is $>$99.7 \%), see Figure \ref{pow_3c279} and Table \ref{top_hits_3c279}.
In the NIR-optical bands the presence of  peaks at $ T_{2}\sim$ 260  d, with
significance  $\geq3\sigma$ in V and K,  stands out.
In R and V bands the adopted  procedure fails in producing a reliable red-noise profile
 due to the prominent flare at  $\sim$ MJD 53700.
  In this case the search for periodicities
was performed  splitting  the light curve in 3 segments, which are overlapping  for the 50\%
of their length \citep[Welch-overlapped-segment-averaging, WOSA, ][]{Welch1967}.
The spectral features are estimated from averaging  the 3 resulting periodograms. 
The obtained   spectra return  peaks at 256 d, which we can associate  with
the one at $\sim 265$ d in K-band (see Figure \ref{pow_3c279} and Table \ref{top_hits_3c279}).

\textit{PKS 1510-089} --
The most significant peak in $\gamma$-ray band is at $T_5$=115 d ($S\sim99\%$), see
 Figure \ref{pow_pks1510}. 
This may be related to the optical peaks at $4\cdot T_5$ and $3\cdot T_5$,  
with $S > 99.0\%$ (see Table \ref{top_hits_pks1510}).
We note that in this source the fit of the red noise continuum may be affected by a number 
of short flares \citep{Sandrinelli2014a}.

\textit{PKS 2005-489} -- 
For this source only REM photometry  is available. 
A conspicuous peak at  $\sim$680-720 d is present in all  the NIR-optical
 bands  (Figure \ref{pow_pks2005} and  Table \ref{top_hits_pks2005}). It may be related to 
 the  peak at   $\sim$ 370 d, which however is supposedly spurious (see above).
 Because of  the length of the light curve ($T_{tot}$=2500 d) the candidate periods  
 between 1000 d and 1500 d   
  are hardly  significant and are not reported in Table \ref{top_hits_pks2005}.
 
\textit{PKS 2155-304} --
We confirm the results presented in \cite{Sandrinelli2014b} about a peak of $ T_3 \sim$ 315 d in
the NIR-optical bands and at $2 \cdot T_3$ in the $\gamma$-rays,
 see Table \ref{top_hits_pks2155} and Figure \ref{pow_pks2155}. 
 In $\gamma$-rays and in K the significances are above $3\sigma$.
 In K it is noticeable also the presence of peaks  with $S \gtrsim99\%$ at 453 d and
 151 d, which with $2 \cdot T_3$ and $T_{3}$ could  be harmonics of the 
 same period $T_{7}=$76 d.


\section{\label{disc} Discussion}
 
The advent of robotic telescopes in the last decade on the one hand, and the 
systematic monitoring of the $\gamma$-ray sky since 2008, gave an unprecedented 
opportunity for exploring  quasi-periodicities in blazars at year-like timescales.
 The presence of possible associated periodicities in both the optical and $\gamma$-ray bands 
 may be indicative of their physical relevance.  
 The most convincing cases are PKS 2155-304, OJ 287 and 3C 279.
 
 In the following we discuss some possible interpretations for the physical mechanisms 
 causing the periodic variability of these sources with a rest-frame year-like duration. 
 As already discussed in \cite{Sandrinelli2014b} and similarly to the
interpretation proposed by \cite{Lehto1996} for OJ 287, 
the observed year-like timescale  periodicity  
could be related to the orbit of a perturbing object. This could
destabilize the accretion flow onto the primary SMBH, modulating the accretion
rate and, as a consequence, the luminosity of the active nucleus. Assuming
the mass of the active SMBH $M_1 \sim 10^9 M_{\odot}$ (typical of blazars) and
that the perturber is significantly less massive ($q=M_2/M_1<0.1$), the
observed periodicity implies a separation between the two bodies of $d \sim
100$ r$_{S}\approx 10^{-2}$ pc, where r$_{S}$ is the Schwarzschild radius of
the central SMBH. Currently available observations do not allow us to
constrain the nature of the perturber. Even assuming that this is a secondary
SMBH of mass $\sim 10^8 M_{\odot}$, the gravitational wave (GW) driven orbital
decay would be of the order of 1 Myr \citep{Peters1964}, making it hard to
observe any frequency drift due to the shrinking of the perturber orbit.

A second possibility is that the observed periodicity is related to the
precession period of the blazar jet. We can draw two distinct scenarios,
depending on whether the jet is forced to be aligned to the SMBH spin or
not. In this second case the jet would be aligned to the angular momentum of
the inner part of the accretion disc. In the case the inner disc is not lying
on the SMBH equatorial plane, it would undergo Lense-Thirring precession
around
the SMBH spin. The Lense Thirring precession period scales as:
\begin{equation}
T_{\rm Prec} = \frac{8 \pi G M_1}{c^3 a} \left(\frac{r}{r_S}\right)^3,
\end{equation}
where $a$ is the SMBH spin parameter. If the accretion flow is geometrically
thick, the inner disk precesses as a solid body, as demonstrated by 
general relativistic magnetohydrodynamic numerical simulations
 \citep[e.g.][]{Fragile2007, Dexter2011}\footnote{We note that a thick geometry is expected for radiatively
  inefficient accretion, making this scenario particularly attractive for BL
  Lacs and specifically for PKS 2155-304 \citep{Ghisellini2011, Sbarrato2012}.}.
    In this case the observed period does provide an estimate of
the size of the inner precessing region. About 900 day correspond to
$\approx 8 r_S$, while $\approx 300$ day
would imply $\approx 5 r_S$ (both the estimates
assume $M_1=10^9 M_{\odot}$ and $a=0.9$).

If instead the jet is aligned with the SMBH spin, the observed periodicity has
to be related to the precession timescale of the spin itself. In order for
the
observed flux to be significantly affected by the precession, the angle
between the SMBH spin and the axis of precession (defined by the total angular
momentum of the system) has to be of the order of the jet opening angle $\sim
1^{\circ}$ or greater. Because of the large mass of the SMBH, the angular
momentum of the accreting gas within the inner $\sim 10 r_S$ cannot cause
such
a large displacement of the SMBH spin. A secondary SMBH orbiting at $\approx 8
r_S$ is required in this case. The secondary mass needed to cause a
significant variation in the observed flux has to be $M_{2} \gtrsim 10^6 M_{\odot}$. 
This last scenario has a number of testable predictions:\\ $(i)$ The
lifetime of such a system would be determined by the GW driven orbital decay
of the binary. The timescale is $\gtrsim 8 \cdot 10^3$ yr for $M_{2} \gtrsim10^6
M_{\odot}$  and it scales with the inverse of $M_2$. For $M_2
\gtrsim 10^7 M_{\odot}$ a drift of the period toward smaller values could be
observable in the next decades;\\ 
$(ii)$ A faster variability, corresponding
to the orbital timescale of the binary, could be observed (with a period of
$\approx 24$ days assuming $q<0.1$).


In a different scenario, 
the  detected quasi-periodic oscillations  in blazars 
could be related to jet emissions.
Variability can be ascribed to
 helical jets or helical structures in jets
\citep[e.g.][]{Camenzind1992} 
 which may be quite  common in  blazars \citep[][]{Villata1999}.
They could arise from hydrodynamical instabilities in magnetized 
jets  \citep{Hardee1999} or  from variations in the jet engine,  e.g.,  accretion 
disc instabilities  \citep[][and reference therein]{Godfrey2012},
also  coupled with the interaction of the jet plasma  with the surrounding medium. 
The  emitting  flow moving around a  helical path 
could produce relatively long-term quasi-periodic changes in Doppler boosted flux \citep{Villata1999}.
In such a picture a rotating helical structure
was proposed   to explain, e.g.,  both the quasi-periodic behavior  \citep[$\sim$8 yr,][]{Raiteri2010}
 of  BL Lacertae  and  the occurrence   and the mean shape of major  radio-optical outbursts 
in AO 0235+16 \citep[$\sim$5.7 yr,][]{Ostorero2004}.
Relativistic shocks \citep[e.g,][]{Marscher1985} exciting jet helical patterns  can be also considered
   \citep[e.g.,][]{Rani2009,Larionov2013}, as well as models based on dominant 
turbulent cells in plasma jet \citep{Marscher1992,Marscher2014}  driving short-lived 
 quasi-periodic oscillations behind a shock \citep[e.g.,][]{Rani2009}.  
The presence of quasi-periodic  peaks  in radio-mapped jet   emissions  are, for instance,
discussed in \cite{Godfrey2012} for the case of  PKS 0637-752,
where  large-scale shocks in continuous flow are invoked
(re-confinement shocks, Kelvin-Helmholtz  instabilities).
The discussion by  \cite{Godfrey2012} proposes   time scales of variability  much larger  
  than those inferred in this paper, but the issue   should be reconsidered in detail.

\section{\label{concl}Conclusions}

The significance of the detected periods is almost never very high, although in 
several case good enough to deserve consideration. As discussed in the text, 
a periodicity analysis of highly irregular and unevenly sampled time-series is 
never an easy task. On the other hand, in some cases the detected periods are
 also connected with periods derived from \textit{Fermi} data, where most of the source 
 of uncertainty do not apply. It is then possible that quasi-periodicities
 of $\sim$ 1 yr in blazars are not rare.
Their origin is not easy to be traced, and detailed models are not yet available.

Progress in the field can come from the study of other blazars, which have 
    both a long optical monitoring, and are relatively bright in the \textit{Fermi} archives. 
    It is also obvious to search for  confirmation of the periodicities in the X-rays where
     large amount of sparse observations are archived. However, the X-rays could be produced 
     independently of the optical and $\gamma$-rays.  
       For the future, observations of the 
       SMARTS and REM type should hopefully be prosecuted for several years,
       largely improving the robustness of the analysis for periodicities of several months / one year
       length.

\acknowledgments

 We are grateful to dr. G.L. Israel for discussions on periodicity search in light curves
affected by red noise. This paper  made use of up-to-date SMARTS optical/near-infrared 
 light curves that are available on 
 line\footnote{\texttt{www.astro.yale.edu/smarts/glast/home.php}}.
We acknowledge the  support of the Italian Ministry of Education  
(grant PRIN-MIUR 2009, 2010, 2011).
This work was also  supported by ASI grant I/004/11/0.

\clearpage

\begin{table*}
\centering
\caption{\label{sample} Blazar sample.}
\begin{tabular}{lcclcl}
\hline\hline             
   Source			&ra\tablenotemark{a}	&dec\tablenotemark{a}& Class\tablenotemark{b}           & SED\tablenotemark{c}		&Redshift\tablenotemark{a}\\
\hline\hline
&&&&&\\
PKS 0537-441		&   05:38:50	&$-$44:05:09	&      BL Lac	&  LSP		&0.896	\\
OJ 287			&   08:54.48	&    +20.06:30	&      BL Lac	& LSP		&0.3060	\\
3C 279			&   12:56:11	&$-$05:47:21  	&      FSRQ	&LSP		&0.536	\\
PKS 1510-089		&   15:12:50	&$-$09:05:59  	&      FSRQ	& LSP		&0.3599	\\
PKS 2005-489		&   20:09:25	&$-$48:49:53	&      BL Lac	& HSP		&0.071	\\
PKS 2155-304		&   21:58:52	&$-$30:13:32	&      BL Lac	& HSP		&0.117	\\
&&&&&\\
\hline	
\end{tabular}
\tablenotetext{a}{From the Simbad archive (\texttt{http://simbad.u-strasg.fr/}).}
\tablenotetext{b}{From \cite{Massaro2012}.}
\tablenotetext{c}{LSP, ISP and HSP  refer to low, intermediate and high synchrotron peaked 
 blazars \citep{Abdo2010}}
 \end{table*}

  \begin{table*}
\caption{\label{3c279lc_tab} REM photometry of 3C 279. 
} 
\centering
\begin{tabular}{cccc}
\hline\hline
Filter		&Time of		&Average 	& Magnitude 	\\	
		&observation	&magnitude	&error		\\
		&[MJD]		&[mag]		&[mag]\\
\hline\hline
&&&\\	
V		&	53467 	&  	 15.61  		&0.05\\
V		&	53469	&  	 15.64 		&0.06\\
V		&	53474	&  	 15.40 		&0.06\\
V		&       53476	&	 15.16		&0.04\\
V		&      	53492	&	 15.80		&0.05\\
....		&  	....		&	....			&  .... \\
&&&\\
\hline
\end{tabular}
\tablecomments{
A full version of  Table \ref{3c279lc_tab} is available
in electronic format.  A portion is 
shown here for guidance regarding its form and content.
}
\end{table*}

\begin{table*}
\caption{\label{3c279prop_tab} Properties of REM NIR-optical light curves of 3C 279. }
\centering
\begin{tabular}{lccccc}
\hline\hline             
Filter	&Mag. range 	& Mean mag.  	&Median	&Flux range			& $\sigma_{rms}$\tablenotemark{a}\\
         	&[mag]		& [mag]		&[mag]	&[mJy]				&[\%]\\	
\hline 
\hline
&&&&&\\
 V 		& 13.15 - 16.83 	 & 15.13 		& 15.14	& 0.73 - 21.77 		& 72 $\pm$	2\\		 
 R 		& 12.70 - 16.58 	 & 14.83 		& 14.95 	& 0.83 - 27.37 		& 83 $\pm$	2\\			 
 I 		& 12.13 - 15.93 	 & 14.27		& 14.32	& 1.08 - 35.68 		& 86 $\pm$	2\\
 J 		& 10.90 - 15.03 	 & 13.04 		& 13.07	& 1.55 - 69.30		& 82 $\pm$	1\\
 H 		& 10.08 - 14.24 	 & 12.20 		& 12.18 	& 2.13 - 97.74 		& 82 $\pm$	1\\
 K 		& 9.58   - 13.30		 & 11.18 		& 11.18	& 3.33 - 99.81 		& 66 $\pm$	1\\
&&&&&\\
\hline
\end{tabular}
\tablenotetext{a}{Fractional variability amplitude.}
\end{table*}

\begin{table*}
\centering
\caption{\label{top_hits_pks0537}  
Most prominent peaks and significances 
($S \gtrsim99\%$) in PKS 0537-441 \textit{observed} power spectra. 
The peaks marked
with (*) appear marginally revealed in our spectral analysis for alias detection.
We note that PKS 0537-441 is not observed by SMARTS in V band.
}	
\begin{tabular}{|c|c|c|c|c|c||c||}													
\hline\hline													
Band	&	  $T_{1}$	&	  $T_{2}$	&	  $T_{3}$	&	  $T_{4}$	&	  $T_{5}$	&	$S \% \gtrsim$	\\
	&	[d]	&	[d]	&	[d]	&	[d]	&	[d]	&		\\
\hline\hline													
R	&	150$\pm$5	&		&		&		&		&	99.0	\\
	&		&	68$\pm$1*	&		&		&		&	99.0	\\
	&		&		&		&	61$\pm$1*	&		&	99.0	\\
	&		&		&		&		&	58$\pm$1*	&	99.0	\\
\hline													
J	&	153$\pm$3	&		&		&		&		&	99.0	\\
	&		&	68$\pm$1*	&		&		&		&	99.0	\\
	&		&		&	64$\pm$1*	&		&		&	99.0	\\
	&		&		&		&	61$\pm$1	&		&	99.7	\\
	&		&		&		&		&	53$\pm$1*	&	99.0	\\
\hline													
K	&	152$\pm$4	&		&		&		&		&	98.0	\\
	&		&	68$\pm$1*	&		&		&		&	99.7	\\
	&		&		&		&	61$\pm$1	&		&	99.7	\\
	&		&		&		&		&	53$\pm$1*	&	99.7	\\
\hline													
\end{tabular}													
\end{table*}

\begin{table*}
\centering
\caption{\label{top_hits_oj}  The same of Table \ref{top_hits_pks0537}  for  OJ 287.
 }											
\begin{tabular}{|c|c|c|c||c||}									
\hline\hline									
Band	&	  $T_{1}$	&	  $T_{2}$	&	  $T_{3}$	&	$S \% \gtrsim$	\\
	&	[d]	&	[d]	&	[d]	&		\\
\hline\hline									
100 MeV-300 Gev	&	412$\pm$25	&		&		&	99.0	\\
\hline									
V	&	435$\pm$24	&		&		&	99.7	\\
\hline									
R	&	436$\pm$27	&		&		&	99.0	\\
\hline									
J	&	436$\pm$25	&		&		&	99.7	\\
	&		&	303$\pm$12*	&		&	99.0	\\
\hline									
K	&	438$\pm$22	&		&		&	99.7	\\
	&		&	296$\pm$10*	&		&	99.7	\\
	&		&		&	203$\pm$5	&	99.7	\\
\hline									
\end{tabular}									
\end{table*}

\begin{table*}
\centering
\caption{\label{top_hits_3c279}   The same of Table \ref{top_hits_pks0537}  for 3C 279. 
 }
\begin{tabular}{|c|c|c|c|c|c|c|c||c||}															
\hline\hline															
Band			&	  $T_{1}$		&	  $T_{2}$	&	  $T_{3}$	&	  $T_{4}$	&	  $T_{5}$	&	  $T_{6}$	&	 $T_{7}$	&	$S\%\gtrsim$	\\
				&	[d]			&	[d]		&	[d]		&	[d]		&	[d]		&	[d]		&	[d]		&			\\
\hline\hline															
100 MeV-300 GeV	&				&				&			&			&	39$\pm$1	&			&			&	99.7	\\
				&				&				&			&			&			&			&	24$\pm$1	&	99.7	\\
\hline															
V				&				&	256$\pm$15	&			&			&			&			&			&	99.7	\\
\hline															
R				&				&	256$\pm$12	&			&			&			&			&			&	99.0	\\
				&				&				&	71$\pm$1	&			&			&			&			&	99.0	\\
				&				&				&			&	66$\pm$1	&			&			&			&	99.0	\\
				&				&				&			&			&			&  29$\pm$1*	&			&	99.0	\\
				&				&				&			&			&			&			&	24$\pm$1	&	99.7	\\
\hline															
K				&	931 $\pm$ 46	&				&			&			&			&			&			&	99.7	\\
				&				&	263$\pm$5	&			&			&			&			&			&	99.7	\\
\hline															
\end{tabular}															
\end{table*}

\begin{table*}
\centering
\caption{\label{top_hits_pks1510}   The same of Table \ref{top_hits_pks0537}  for PKS 1510-089.}
\begin{tabular}{|c|c|c|c|c|c|c|c||c||}													
\hline\hline													
Band	&	  $T_{1}$	&	  $T_{2}$	&	  $T_{3}$	&	  $T_{4}$	&	  $T_{6}$	&	$S \% \gtrsim$	\\
\hline\hline													
100 MeV-300 Gev	&		&		&		&		&	115$\pm$5	&	99	\\
\hline													
V	&	490$\pm$34	&		&		&		&		&	99.0	\\
	&		&	325$\pm$16	&		&		&		&	99.7	\\
\hline													
R	&	490$\pm$37	&		&		&		&		&	99.7	\\
	&		&	325$\pm$13	&		&		&		&	99.0	\\
	&		&		&		&	206$\pm$9	&		&	99.7	\\
\hline													
J	&	474$\pm$36	&		&		&		&		&	99.7	\\
	&		&	321$\pm$15	&		&		&		&	99.0	\\
	&		&		&		&	203$\pm$9	&		&	99.7	\\
\hline													
K	&	474$\pm$34	&		&		&		&		&	99.7	\\
	&		&	321$\pm$16	&		&		&		&	99.0	\\
	&		&		&	262$\pm$10*	&		&		&	99.0	\\
	&		&		&		&	207$\pm$10	&		&	99.7	\\
\hline													
\end{tabular}													
\end{table*}

\begin{table*}
\centering
\caption{\label{top_hits_pks2005}   The same of Table \ref{top_hits_pks0537}  for PKS 2005-489.  
}
\begin{tabular}{|c|c|c|c|c||c||}									
\hline\hline									
Band	&	  $T_{1}$ 	&	  $T_{2}$ 	&	  $T_{3}$	&	$S\%\gtrsim $	\\
	&	[d]	&	[d]	&	[d]	&		\\
\hline\hline									
V	&	719$\pm$64*	&		&		&	97	\\
	&		&	360$\pm$44*	&		&	99	\\
	&		&		&	93$\pm$2*	&	99.7	\\
\hline									
R	&	693$\pm$74*	&		&		&	99	\\
	&		&	381$\pm$48*	&		&	99.7	\\
	&		&		&	93$\pm$1*	&	99	\\
\hline									
J	&	683$\pm$51*	&		&		&	99	\\
	&		&	381$\pm$39*	&		&	99.7	\\
\hline									
K	&	722$\pm$47	&		&		&	99	\\
	&		&	361$\pm$28*	&		&	99	\\
\hline									
\end{tabular}									
\end{table*}

\begin{table*}
\centering
\small
\caption{\label{top_hits_pks2155}   The same of Table \ref{top_hits_pks0537}  for PKS 2155-304.  }
\begin{tabular}{|c|c|c|c|c|c|c|c|c|c||c||}																					
\hline\hline																					
Band	&	  $T_{1}$	&	  $T_{2}$	&	  $T_{3}$	&	  $T_{4}$	&	  $T_{5}$	&	  $T_{6}$	&	  $T_{7}$	&	  $T_{8}$	&	  $T_{9}$	&	$S \% \gtrsim$	\\
	&	[d]	&	[d]	&	[d]	&	[d]	&	[d]	&	[d]	&	[d]	&	[d]	&	[d]	&		\\
\hline\hline																					
100 MeV-300 GeV	&	642$\pm$59	&		&		&		&		&		&		&		&		&	99.7	\\
	&		&		&		&		&		&		&		&	61$\pm$1*	&		&	99.0	\\
	&		&		&		&		&		&		&		&		&	52$\pm$1	&	99.0	\\
\hline																					
V	&		&		&	318$\pm$14	&		&		&		&		&		&		&	99.0	\\
	&		&		&		&		&		&		&	75$\pm$1		&		&		&	99.7	\\
	&		&		&		&		&		&		&		&	64$\pm$1*	&		&	99.0	\\
\hline																					
R	&		&		&	318$\pm$14	&		&		&		&		&		&		&	99.0	\\
\hline																					
J	&		&		&	310$\pm$15	&		&		&		&		&		&		&	99.0	\\
	&		&		&		&		&		&		&	76$\pm$1		&		&		&	99.0	\\
	&		&		&		&		&		&		&		&	63$\pm$1*	&		&	99.0	\\
\hline																					
K	&		&	453$\pm$16	&		&		&		&		&		&		&		&	99.7	\\
	&		&		&	311$\pm$14	&		&		&		&		&		&		&	99.7	\\
	&		&		&		&	166$\pm$2	&		&		&		&		&		&	99.7	\\
	&		&		&		&		&	151$\pm$2	&		&		&		&		&	99.7	\\
	&		&		&		&		&		&	93$\pm$1*	&		&		&		&	99.0	\\
	&		&		&		&		&		&		&	76$\pm$1		&		&		&	99.0	\\
\hline																					
\end{tabular}																					
\end{table*} 																					
\clearpage

\clearpage

\begin{thebibliography}{}
 

\bibitem[Abdo et al.(2010)]{Abdo2010} Abdo, A.~A., Ackermann, 
M., Ajello, M., et al.\ 2010, \apj, 722, 520 

\bibitem[\protect\citeauthoryear{Bonning et 
al.}{2012}]{Bonning2012} Bonning E., et al., 2012, ApJ, 756, 13 
%
\bibitem[Camenzind 
\& Krockenberger(1992)]{Camenzind1992} Camenzind, M., \& Krockenberger, M.\ 1992, \aap, 255, 59 

\bibitem[Covino et al.(2004)]{Covino2004} Covino, S., Stefanon, 
M., Sciuto, G., et al.\ 2004, \procspie, 5492, 1613 

\bibitem[\protect\citeauthoryear{Deeming}{1975}]{Deeming1975} Deeming T.~J., 1975, Ap\&SS, 36, 137 

\bibitem[\protect\citeauthoryear{Dexter \& Fragile}{2011}]{Dexter2011}
Dexter J., Fragile P.~C., 2011, ApJ, 730, 36

\bibitem[\protect\citeauthoryear{Edelson et 
al.}{2002}]{Edelson2002} Edelson R., Turner T.~J., Pounds K., 
Vaughan S., Markowitz A., Marshall H., Dobbie P., Warwick R., 2002, ApJ, 
568, 610 

\bibitem[\protect\citeauthoryear{Falomo, Pian, 
\& Treves}{2014}]{Falomo2014} Falomo R., Pian E., Treves A., 2014, A\&ARv, 22, 73 

\bibitem[Fragile et al. (2007)]{Fragile2007} Fragile P.~C., Blaes O.~M.,
Anninos P., Salmonson J.~D., 2007, ApJ, 668, 417

\bibitem[Gab{\'a}nyi et 
al.(2007)]{Gabanyi2007} Gab{\'a}nyi, K.~{\'E}., Marchili, N., Krichbaum, T.~P., et al.\ 2007, \aap, 470, 83 

\bibitem[Ghisellini et al. (2011)]{Ghisellini2011} Ghisellini G.,
  Tavecchio F., Foschini L., Ghirlanda G., 2011, MNRAS, 414, 2674

\bibitem[Ghisellini(2013)]{Ghisellini2013} Ghisellini, G.\ 2013, 
Lecture Notes in Physics, Radiative Processes in High Energy Astrophysics,
Berlin Springer Verlag, 873 

\bibitem[Godfrey et al.(2012)]{Godfrey2012} Godfrey, L.~E.~H., 
Lovell, J.~E.~J., Burke-Spolaor, S., et al.\ 2012, \apjl, 758, L27 

\bibitem[\protect\citeauthoryear{Gonz{\'a}lez-P{\'e}rez, Kidger, 
\& Mart{\'{\i}}n-Luis}{2001}]{Gonzalez2001}  
Gonz{\'a}lez-P{\'e}rez J.~N., Kidger M.~R., Mart{\'{\i}}n-Luis F., 2001, AJ, 122, 2055 

\bibitem[Graham et al.(2015)]{Graham2015} Graham, M.~J., 
Djorgovski, S.~G., Stern, D., et al.\ 2015, \nat, 518, 74 

\bibitem[\protect\citeauthoryear{Hardee 
\& Rosen}{1999}]{Hardee1999} Hardee P.~E., Rosen A., 1999, ApJ, 524, 650 

\bibitem[\protect\citeauthoryear{Hasselmann}{1976}]{Hasselmann1976}
 Hasselmann K., Theory Tellus,1976, 28 (6), 473 

\bibitem[\protect\citeauthoryear{Hudec et 
al.}{2013}]{Hudec2013} Hudec R., Ba{\v s}ta M., Pihajoki P., Valtonen M., 2013, A\&A, 559, A20 

\bibitem[Israel 
\& Stella(1996)]{Israel1996} Israel, G.~L., \& Stella, L.\ 1996, \apj, 468, 369 

\bibitem[Kidger(2000)]{Kidger2000} Kidger, M.~R.\ 2000, \aj, 119, 
2053 

\bibitem[\protect\citeauthoryear{Larionov et 
al.}{2013}]{Larionov2013} Larionov V.~M., et al., 2013, ApJ, 768, 40 

\bibitem[Lehto 
\& Valtonen(1996)]{Lehto1996} Lehto, H.~J., \& Valtonen, M.~J.\ 1996, \apj, 460, 207 

\bibitem[Li et al.(2015)]{Li2015} Li, H.~Z., Chen, L.~E., Yi, 
T.~F., et al.\ 2015, \pasp, 127, 1 

\bibitem[Marscher 
\& Gear(1985)]{Marscher1985} Marscher, A.~P., \& Gear, W.~K.\ 1985, \apj, 298, 114

\bibitem[\protect\citeauthoryear{Marscher}{1992}]{Marscher1992} Marscher, A. P., Gear, W. K., \& Travis, J. P. 1992, in Variability of Blazars,
ed. E. Valtaoja \& M. Valtonen (Cambridge Univ. Press), 85

\bibitem[\protect\citeauthoryear{Marscher}{2014}]{Marscher2014} 
Marscher A.~P., 2014, ApJ, 780, 87 

\bibitem[Massaro et al. (2012)]{Massaro2012} Massaro, E., Giommi, P., Leto, C.;, et  al\ 2012, 
  Multifrequency Catalogue of Blazars (3rd Edition) 
 (ARACNE Ed.) 

\bibitem[Mattox et al.(1996)]{Mattox1996} Mattox, J.~R., Bertsch, 
D.~L., Chiang, J., et al.\ 1996, \apj, 461, 396 

\bibitem[\protect\citeauthoryear{Nandra et al.}{1997}]{Nandra1997} 
Nandra K., George I.~M., Mushotzky R.~F., Turner T.~J., Yaqoob T., 1997, 
ApJ, 476, 70 

\bibitem[Ostorero et 
al.(2004)]{Ostorero2004} Ostorero, L., Villata, M., \& Raiteri, C.~M.\ 2004, \aap, 419, 913 

\bibitem[\protect\citeauthoryear{Peters}{1964}]{Peters1964}
Peters P.C., 1964, Phys. Rev. B, 136, 1224

\bibitem[Press(1978)]{Press1978} Press, W.~H.\ 1978, Comments on 
Astrophysics, 7, 103 

\bibitem[Raiteri et 
al.(2001)]{Raiteri2001} Raiteri, C.~M., Villata, M., Aller, H.~D., et al.\ 2001, \aap, 377, 396 

\bibitem[Raiteri et 
al.(2010)]{Raiteri2010} Raiteri, C.~M., Villata, M., Bruschini, L., et al.\ 2010, \aap, 524, A43 

\bibitem[Rani et al.(2009)]{Rani2009} Rani, B., Wiita, P.~J., 
\& Gupta, A.~C.\ 2009, \apj, 696, 2170 

\bibitem[Sandrinelli et 
al.(2014a)]{Sandrinelli2014a} Sandrinelli, A., Covino, S., \& Treves, A.\ 2014a, \aap, 562, A79 

\bibitem[\protect\citeauthoryear{Sandrinelli et al.}{2014b}]
{Sandrinelli2014b} Sandrinelli A., Covino S., Treves A., 2014b, ApJ, 793, L1 

\bibitem[\protect\citeauthoryear{Sbarrato et al.}{2012}]{Sbarrato2012}
Sbarrato T., Ghisellini G., Maraschi L., Colpi M., 2012, MNRAS, 421, 1764

\bibitem[Scargle(1982)]{Scargle1982} Scargle, J.~D.\ 1982, \apj, 
263, 835 

\bibitem[\protect\citeauthoryear{Schulz \& Statteger}{1997}]{Schulz1997}
 Schulz, M. \& Statteger, K., 1997, Comput. Geosci., 23, 929 

\bibitem[\protect\citeauthoryear{Schulz \& Mudelsee}{2002}]{Schulz2002}
 Schulz, M. \& Mudelsee, M., 2002, Comput. Geosci., 28(3), 421

\bibitem[\protect\citeauthoryear{Schwarzenberg-Czerny}{1991}]{Schwar1991}
 Schwarzenberg-Czerny A., 1991, MNRAS, 253, 198 

\bibitem[Sillanp\"a\"a et al.(1988)]{Sillanpaa1988} Sillanp\"a\"a, A., 
Haarala, S., Valtonen, M.~J., Sundelius, B., 
\& Byrd, G.~G.\ 1988, \apj, 325, 628 

\bibitem[\protect\citeauthoryear{Skrutskie et 
al.}{2006}]{Skrutskie2006} Skrutskie M.~F., et al., 2006, AJ, 131, 
1163 

\bibitem[Sundelius et al.(1997)]{Sundelius1997} Sundelius, B., Wahde, 
M., Lehto, H.~J., \& Valtonen, M.~J.\ 1997, \apj, 484, 180 

\bibitem[\protect\citeauthoryear{Timmer 
\& Koenig}{1995}]{Timmer1995} Timmer J., Koenig M., 1995, A\&A, 300, 707 

\bibitem[van der Klis(1989a)]{vanderKlis1989a} van der Klis, M.\ 1989a, 
NATO Advanced Science Institutes (ASI) Series C, 262, 27

\bibitem[Valtonen 
\& Ciprini(2012)]{Valtonen2012} Valtonen, M., \& Ciprini, S.\ 2012, \memsai, 83, 219 

\bibitem[\protect\citeauthoryear{van der 
Klis}{1989b}]{vanderKlis1989b} van der Klis M., 1989b, ARA\&A, 27, 517 

\bibitem[Vaughan(2005)]{Vaughan2005} Vaughan, S.\ 2005, \aap, 431, 391 

\bibitem[Vaughan(2010)]{Vaughan2010} Vaughan, S.\ 2010, \mnras, 
402, 307 

\bibitem[Villata 
\& Raiteri(1999)]{Villata1999} Villata, M., \& Raiteri, C.~M.\ 1999, \aap, 347, 30 

\bibitem[\protect\citeauthoryear{Welch}{1967}]{Welch1967} 
Welch, P.D.,  1967, IEEE Transactions on Audio and Electroacoustics
15 (2), 70 

\bibitem[Zerbi et al.(2004)]{Zerbi2004} Zerbi, F.~M., Chincarini, 
G., Ghisellini, G., et al.\ 2004, \procspie, 5492, 1590 

\bibitem[\protect\citeauthoryear{Zhang et al.}{2014}]{Zhang2014} 
Zhang B.-K., Zhao X.-Y., Wang C.-X., Dai B.-Z., 2014, RAA, 14, 933 

\bibitem[\protect\citeauthoryear{Zhou 
\& Sornette}{2002}]{Zhou2002} Zhou W.-X., Sornette D., 2002, IJMPC, 13, 137 


\end{thebibliography}
\end{document}